\documentclass[useAMS,usenatbib]{mn2e}
\usepackage[T1]{fontenc}
\usepackage{ae,aecompl}
\usepackage[modulo, switch]{lineno} 
\usepackage{amssymb}
\usepackage{graphicx}
\usepackage{epstopdf}
\usepackage{amsmath}    
\usepackage{amssymb}    
\usepackage{multicol}        
\usepackage{bm}     
\usepackage{pdflscape}  

\title[Modelling Kepler Red Giants in Eclipsing Binaries]
{Modelling Kepler Red Giants in Eclipsing Binaries: Calibrating the Mixing-Length Parameter with Asteroseismology}
\author[Tanda Li et al.]{Tanda Li$^{1,2,3}$\thanks{E-mail:
tanda.li@sydney.edu.au},
Timothy R. Bedding$^{1,2}$, 
Daniel Huber$^{4,1,5,2}$, 
Warrick H. Ball$^{7,8,9,2}$,
\newauthor
Dennis Stello$^{1,2,6}$,
Simon J. Murphy$^{1,2}$,
Joss Bland-Hawthorn$^{1}$
\\
$^{1}$Sydney Institute for Astronomy (SIfA), School of Physics, University of Sydney, NSW 2006, Australia\\
$^{2}$Stellar Astrophysics Centre, Department of Physics and Astronomy, Aarhus University, Ny Munkegade 120, DK-8000 Aarhus C, Denmark\\
$^{3}$Key Laboratory of Solar Activity, National Astronomical Observatories, Chinese Academy of Science, Beijing 100012, China \\
$^{4}$Institute for Astronomy, University of Hawai'i, 2680 Woodlawn Drive, Honolulu, HI 96822, USA\\
$^{5}$SETI Institute, 189 Bernardo Avenue, Mountain View, CA 94043, USA\\
$^{6}$School of Physics, University of New South Wales, Australia\\
$^{7}$Institut f\"ur Astrophysik, Georg-August-Universit\"at G\"ottingen, Friedrich-Hund-Platz 1, 37077 G\"ottingen, Germany\\
$^{8}$Max-Planck-Institut f\"ur Sonnensystemforschung, Justus-von-Liebig-Weg 3, 37077 G\"ottingen, Germany\\
$^{9}$School of Physics and Astronomy, University of Birmingham, Edgbaston, Birmingham, B15 2TT, UK \\
}

\begin{document}
%
\date{}

\pagerange{\pageref{firstpage}--\pageref{lastpage}} \pubyear{}

\maketitle

\label{firstpage}

\begin{abstract}
Stellar models rely on a number of free parameters.
High-quality observations of eclipsing binary stars
observed by $Kepler$ offer a great opportunity to calibrate model parameters for evolved stars. 
Our study focuses on six $Kepler$ red giants with the goal of calibrating the mixing-length parameter of convection 
as well as the asteroseismic surface term in models. 
We introduce a new method to improve the identification of oscillation
modes which exploits theoretical frequencies to guide the mode identification (`peak-bagging')
stage of the data analysis. 
Our results indicate that the convective mixing-length parameter ($\alpha$) is $\approx$14\% larger for red giants than for the Sun, in 
agreement with recent results from modelling the APOGEE stars.
We found that the asteroseismic surface term (i.e. the frequency offset between the observed and predicted modes) correlates with stellar parameters ($T_{\rm{eff}}$, $\log g$) and the mixing-length parameter.
This frequency offset generally decreases as giants evolve.
The two coefficients $a_{-1}$ and $a_3$ for the inverse and cubic terms
that have been used to describe the surface term correction are found to correlate linearly. 
The effect of the surface term is also seen in the p-g mixed 
modes, however, established 
methods for correcting the effect are not able to properly correct
the g-dominated modes in late evolved stars.
 \end{abstract}


\section{Introduction}

Stellar models describe internal structures and evolutionary states
of stars. The basic equations were established decades ago and are
able to reproduce the general features of
stars. However, current theoretical models are commonly working with
a number of free parameters.
The atmospheres of stars can be measured by photometry and spectroscopy,
while the stellar masses, ages, and internal structures are mostly provided by models.
Thus, both observational and theoretical calibrations
are required for a proper understanding of stars.
Free parameters, however, undermine the reliability
of theoretical models and increase the true uncertainties of 
modelling results. These parameters need calibrations
for different types of stars, but unfortunately, theoretical 
tests beyond the solar case are sparse, especially for red giants.  

Among $Kepler$ red giants, a few have been identified as
eclipsing binaries (EBs). Combining eclipses with radial velocities from spectra
allows masses and radii of both companions of the binary to be determined from 
dynamical modelling \citep{frandsen13,PG16}, providing the key 
constraints for stars. The structures and evolutionary histories of 
detached companions are similar to that of single stars, and hence they are a good 
population for testing the model parameters. 
Red giants have helium cores and 
burn hydrogen in a surrounding shell. Characteristics of the core decide
the temperature of the H-burning shell and hence determine the 
total luminosity. Tight constraints on the physics of the 
models for red giants would require measurements of the 
core \citep{Lagarde15}. 
Modelling the red giants 
with precise measurements of the solar-like oscillations
by $Kepler$ can provide powerful
constraints on stellar properties \citep{perez16, jiang11}.
The acoustic modes mainly 
probe the envelope, but the mixed modes caused by 
the p and g mode coupling probe the properties of the core 
\citep{Tim11, Mosser14, Lagarde16}. Thus, red giants in binary 
systems with well-defined oscillating patterns, 
offer an important opportunity to test free parameters in stellar models.

Low-mass dwarfs and red giants have 
convective envelopes, where the energy transport is dominated 
by convection. Simulating the real dynamic process for the whole region
comes with great difficulties in computation, hence a mixing-length
approximation is adopted in stellar evolution models. The basic
idea of the mixing-length theory is to define a characteristic length for a fluid
parcel, over which it maintains its original properties before mixing with the surrounding fluid. 
The mixing-length parameter ($\alpha \equiv l/H_{p}$) characterizes this length in the stellar model,
where $l$ is the mixing length, and $H_{p}$ is the pressure
scale height. For low-mass red giants, 
the value of $\alpha$ mainly correlates with 
the total radius and the structure of the envelope.
With the independent measurements of masses and radii, as
well as the internal structures probed by asteroseismology,  
this parameter could be constrained properly by seismic 
red giants in eclipsing binaries.

The surface term is the systematic offset of individual
frequencies between models and observations and arises from poor
modelling of the near-surface layers \citep{CD88,CD96,CDT97}. 
\citet{RV03} suggested using the 
ratio of small to large separations to fit models to observations.
The advantage of this method is avoiding the uncertainty
of the outer layers since the ratio is mainly determined by the interior structure.
\citet{kje08} fitted oscillation
frequencies of the Sun and three well identified sun-like dwarfs with the surface
term corrected by a power law. The approach showed good agreement
and has been widely applied to other stars. 
A later update of the surface correction formulas given 
by \citet{ball14} took mode inertia into account
and modelled the surface effects by terms
proportional to $\nu^{-1}/I$ and $\nu^{3}/I$ based on the discussion
of potential asymptotic forms for frequency shifts by
\citet{Gough90}. \citet{sonoi15} further applied  
3D hydrodynamical simulations to derive
expressions for the surface correction. 
The surface term is strongly correlated to surface properties, 
such as effective temperature and surface gravity.
It also relates to model parameters, especially the mixing length 
parameter, which determine the structure of the envelope.
For this reason, the surface term calibrated to the Sun
does not apply directly to other stars and all current methods 
introduce free parameters because of the uncertainty of the surface layers. The seismic red giants in eclipsing binaries, 
as mentioned above, provide us some of the best constrained 
stars apart from the Sun and they can be the ideal sample
for studying the surface term in evolved stars.

In this work, we selected six $Kepler$ red giants 
in eclipsing binaries with detectable solar-like 
oscillations from \citet{PG16} for
calibrating the mixing-length parameter and the 
surface term. We used four-years of $Kepler$ data to get seismic frequencies
and then generate theoretical models for each star.
All available observed constraints, namely, mass, radius, 
atmospheric parameters and stellar oscillations, are used for 
the calibrations.  

\section{Kepler EB Targets}

\subsection{Target Selection}
In this work we used red giants in detached eclipsing binaries from
\citet{PG16}, who used photometric data from the $Kepler$
mission and spectra obtained by the
3.5-m ARC telescope at APO(ARCES) and the
Sloan Digital Sky Survey(APOGEE) for measuring eclipses 
and radial velocities of the binaries.
These binary systems each comprise one dwarf and one giant ranging
from 1.0 to 1.6$M_{\odot}$. The giants are slightly more massive 
than their dwarf companions. The companions are sufficiently distant 
for them to evolve independently, because 
the separations are roughly 10 to 20 times the radii of the primary stars.
From the sample, we selected six red giants 
showing high S/N solar-like oscillations  
as our targets.
Fundamental stellar parameters of the selected stars
are summarised in Table 1. 
Obvious systematic differences can be seen in the effective temperatures 
and metallicities between ARCES and APOGEE. 
Because both observations were based on
high-resolution spectra, we used their average $T_{\rm{eff}}$ and [Fe/H] 
and the mean uncertainties of the two parameters in this work, 
which were also given in Table 1.

\begin{table*}
 \centering
 \begin{minipage}{160mm}
  \caption{Observations of the Six $Kepler$ Eclipsing Binaries}
  \begin{tabular}{@{}llllllllll@{}}
  \hline
   Star&\multicolumn{3}{c}{Spectroscopic parameters} && \multicolumn{3}{c}{Dynamical modelling \footnote{\citet{PG16}}} & \multicolumn{2}{c}{Asteroseismology \footnote{This work}} \\

       & $T_{\rm{eff}} $  & $\log g$ & [Fe/H]  & Ref. & $M$ & $R$ & $\log g$ & $\Delta \nu$&$\nu _{\rm{max}}$\\
       & [$K$] & [dex] & [dex] & & [$M_{\odot}$] & [$R_{\odot}$] & [dex] & [$\mu$Hz] & [$\mu$Hz] \\
 \hline
  KIC 4663623 & 4812(92) & 2.7(2) & -0.13(06) & ARCES  &1.4(1)&9.8(3)&2.60(3)&5.18(1) &53.5(7) \\
          & 4803(91) & 2.7(1) & 0.16(04) & APOGEE &&&&& \\
          & 4808(92)& - & 0.01(05) & Adopted &&&&& \\
 \hline
  KIC 5786154 & 4747(100) & 2.6(2) & -0.06(06) & ARCES &1.06(6)&11.4(2)&2.35(2)&3.51(1) & 30.1(4)\\
          & 4747(100)& - & -0.06(06) & Adopted &&&&& \\
\hline
  KIC 7037405 & 4516(36) & 2.5(2) & -0.34(06) & ARCES &1.25(4)&14.1(2)&2.24(1)&2.78(1) &22.2(7)\\
          & 4542(91) & 2.3(1) & -0.13(06) & APOGEE &&&& &\\
          & 4529(64)& - & -0.24(06) & Adopted &&&&& \\
\hline
  KIC 8410637 &4699(91) & 2.7(1)& 0.16(03) & APOGEE&1.56(3)&10.7(1)&2.57(1)&4.63(1) &46.3(9) \\
          &4800(80)&2.8(2)&0.08(13)&\citet{frandsen13}&&&&&\\
          & 4750(86)& - & 0.12(08) & Adopted &&&&& \\
\hline
 KIC 9540226 & 4692(65) & 2.2(2) & -0.33(04) &ARCES&1.33(5)&12.8(1)&2.349(8)&3.19(1)&27.8(4)\\
         & 4662(91) & 2.5(1) &-0.16(08) & APOGEE&&&&&\\
         & 4677(78)& - & -0.25(06) & Adopted &&&&& \\
\hline
 KIC 9970396 &4916(68) & 3.1(1) & -0.23(03) & ARCES &1.14(3)&8.0(2)&2.69(2)&6.30(1) &63.8(5)\\
         &4789(91) & 2.7(1) & -0.18(07) & APOGEE&&&&&\\
         & 4853(80)& - & -0.20(05) & Adopted &&&&& \\
  \hline
\end{tabular}

{Note: The first column gives the KIC number; the second to 
fourth columns include atmospheric parameters; the sixth to eighth columns show
the mass, radius and surface gravity obtained by \citet{PG16} from
dynamical modelling; and the last two columns list the asteroseismic observables 
($\Delta\nu$ and $\nu _{\rm{max}}$) extracted by the SYD Pipeline \citep{huber09}.}

\end{minipage}
\end{table*}

\subsection{Stellar Oscillation and Data Analysis}

We used $Kepler$ long-cadence data and
the SYD pipeline \citep{huber09} to 
extract solar-like oscillations of the six red giants.
The reduction steps for preparing light curves include cutting out safe-modes,
correcting the jumps, removing long-period variation by a high-pass filter
and the eclipses using the orbital periods in \citet{PG16}.
The top panel in Figure 1 includes the normalised light curves
of KIC 9970396 after the reduction.  
We then calculated power spectra, estimated the region of power excess, 
fitted to and corrected for the background on power spectra.  
The bottom panel in Figure 1 shows 
the background-corrected power spectra of KIC 9970396. 
It should be noted that the unit S/N indicates the ratio between 
the power density of stellar oscillations and the power density of 
the background.

Because the long-term variations (from eclipses, instrumental drifts, and 
stellar spots) and the granulation background had been removed
from the power spectra we used to study stellar oscillation, 
the noise is mainly white.
The power spectrum of white noise follows a chi-square  
distribution with two degrees of freedom in the frequency domain
\citep{chaplin2002}.
We used the frequency bins from 5 to 290 $\mu$Hz to  
calculate the probability density function (PDF) of the power density,
which can be seen in the bottom right panel in Figure 1.
The larger the power excess over the white noise distribution, 
the more likely that power comes from stellar signal.
For the PDF of a given spectrum, a cumulative probability 
can be set up to separate the signal and 
the noise. And we used a 95-percent threshold.
To evaluate the probability
of each frequency bin being signal statistically, 
we applied Monte-Carlo simulations and produced 
1,000 simulated power spectra by multiplying the 
power spectrum by a random noise distribution
following $\chi^{2}$ with 2 degrees of freedom.
For each simulated power spectrum, we used the 
95-percent threshold and marked 
every frequency bin by a flag of 'Signal' or 'Noise'. 
Lastly, each frequency bin on a power spectrum 
had 1,000 flags. And the percentage of the 'Signal' flags
was used to describe its signal probability 
($\mathcal{P}_{\rm{signal}}$ hereafter). 
In the bottom left panel in Figure 1,
the colour code shows $\mathcal{P}_{\rm{signal}}$ 
of the frequency bins in $\nu_{\rm{max}} \pm 5 \Delta\nu$ of KIC 9970396.
After the above analysis, the weight
of each frequency bin in the following 
process could be evaluated by its 
$\mathcal{P}_{\rm{signal}}$.


\begin{figure*}
 \includegraphics{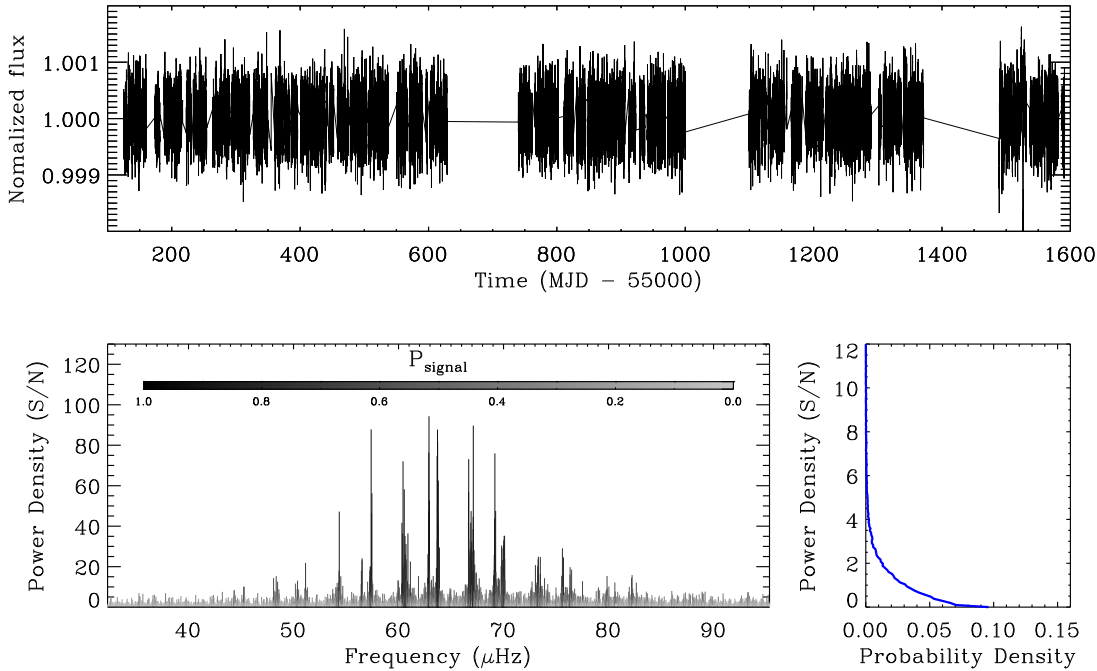}
 \caption{Top: the normalised flux of the $Kepler$ light curve of KIC 9970396.
 Bottom left: background-corrected power spectrum of solar-like oscillations of KIC 9970396 generated by SYD pipeline \citep{huber09}. Colour code indicates the probability of each frequency bin being signal ($\mathcal{P}_{\rm{signal}}$). 
 Bottom right: the probability distribution of the power density. }
\label{fig:ts-ps}
\end{figure*}

Peak-bagging of individual frequencies was then
carried out. Low-degree p modes are essentially
equidistant in frequency, so for a given degree ($l$) with a separation 
of $\Delta\nu$, the mode frequency is 
approximated by 
\begin{equation}
 \nu_{nl} \approx \Delta \nu(n + \frac{l}{2} + \epsilon) - \delta_{nl}, 
\end{equation}
where $n$ is the radial order, $\epsilon$ is a parameter  
related to stellar surface features, and $\delta_{nl}$ is the small
separation. The large separation for the stars were calculated by 
the SYD pipeline. And the small separation are roughly measured by hand  
in the \'echelle diagram. We got the first guesses of the modes
for $l$ = 0 and 2 by this function.
The $l$ = 1 modes couple to g-modes more strongly, and
hence multiple modes were seen for a given order ($n$) in
the power spectrum. 
We measured the median frequency of the multiple $l$ = 1
peaks as the initial guess for the most-p-like mode.
For a proper estimation
of individual modes, we applied the same Monte-Carlo simulations
as mentioned above and generated 1,000 simulated power spectra 
for each star. 
The least-squares method was then used to fit the individual mode 
in each simulated power spectrum.
A Lorentzian function was used for fitting the modes for $l$ = 0 and 2. 
For fitting the most-p-like modes for $l$ = 1, we used a Gaussian function
to fit the shape of the power excess. 
The Lorentzian/Gaussian function had three free parameters, namely, 
the centre, the amplitude, and the width. We fixed the baseline 
as 1.0, which is the median value of the noise in the unit of S/N.
The frequency range for fitting $l$ = 0 and 2 modes was 
set as $\pm 0.6 \delta \nu_{02}$ centred at the
initial guesses. The fitting region for $l$ = 1 modes 
varied for different cases and was determined through visual inspection.
We also adopted the $\mathcal{P}_{\rm{signal}}$ as 
the weight of every frequency bin when calculating the least-squares
of a fitting. First, signal deserves larger weight than noise. 
Secondly, this fitting method could get reasonable width for 
the radial modes with low S/N. For some cases when the power excess
of a mode shows multiple peaks in the power spectra (e.g. the $l$ = 0 mode shown in the
left bottom panel in Figure 2),
the least-squares method without taking account the weight
may fit one of these peaks instead of fitting all of them, giving
an unrealistically narrow width for the mode.  

\begin{figure}
  \includegraphics[width=\columnwidth]{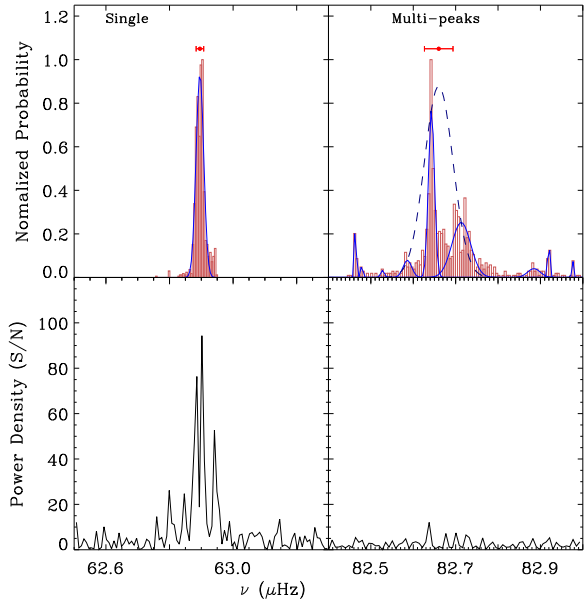}
 \caption{Top: Two typical cases of probability distributions 
 of individual $l$ = 0 modes based 
 on 1,000 Monte-Carlo simulations. 
 left: single peak (High S/N ratio); 
 right: multiple peaks (low S/N ratio). Blue solid lines 
 indicate the Gaussian profile for fitting the distributions. 
 Red error
 bars in upper panels are estimates of the mode frequencies and their uncertainties.
 Bottom: original power spectrum corresponding to the two cases.
 }
\label{fig:mc_nu}
\end{figure}

After the fitting process above, each mode were measured 
(the centre of the Lorentzian profile) on the 1,000 simulated power spectra.
Then a probability distribution of these frequencies could be 
obtained for estimating a mode. 
Two typical examples of the probability distribution
are given in Fig. \ref{fig:mc_nu}.  
For a mode with high S/N (bottom left panel), 
a single clear distribution can be obtained (top left panel). 
We fitted this probability distribution with a 
Gaussian profile then adopted the centre and 1-sigma deviation 
as the estimate of the mode frequency (shown by the red error bar). 
At low S/N (bottom right panel), 
a mode tends to get multiple solutions (top right panel). 
For this case, we first fitted each of these solutions
individually by a Gaussian (blue solid lines)
and then fitted the centres of these solutions
with another Gaussian function (the blue dashed line) to
for getting the frequency and its uncertainty 
(the red error bar).
The method above was used to determine the radial modes ($l$ = 0) 
and the most p-like modes for $l$ = 1 and 2. The identification 
of individual mixed modes for $l$ = 1 will be discussed in Section 4.

\section[]{Theoretical Models}
\subsection{Stellar Models and Input Physics}
In this work, we used Modules for Experiments in Stellar Astrophysics
(MESA, version 8118) to compute stellar evolutionary tracks and generate
structural models. MESA is an open-source stellar evolution package
that is undergoing active development. Detailed descriptions
can be found in \citet{Paxton11,Paxton13, Paxton15}.

We adopted the input physics of
the calibrated solar model of the 'test\_suite' case except for
the atomic diffusion.
In summary, the solar chemical mixture [$(Z/X)_{\odot}$ = 0.0229]
provided by \citet{GS98} was adopted because 
solar models calibrated with this mixture \citep{Bi11} fit the internal
structures from helioseismic inversion better than
those with more recent measurements \citep{AGS05,ASP09}. To
determine initial abundances of hydrogen and helium
($X_{\rm{init}}$ and $Y_{\rm{init}}$) for a given content of heavy
elements ($Z_{\rm{init}}$), we used the following formula: 
\begin{equation}
 {Y_0} = 0.249 \\
\end{equation}
\begin{equation}
 {Y_{\rm{init}}} = {Y_0} + \frac{{\Delta Y}}{{\Delta Z}}{Z_{\rm{init}}} \\
 \end{equation}
 \begin{equation}
 X + Y + Z = 1. \\
\end{equation}
The primordial helium abundance ($Y_0$) is determined by \citet{PC16} 
using the Planck power spectra, Planck lensing, and some external
data such as baryonic acoustic oscillations. The
ratio ${\Delta Y}/{\Delta Z}$ in Eq. 3 can be computed by the initial
abundances of helium and heavy elements of the Sun. We adopted the $Y_{\rm{\odot,init}}$
and $Z_{\rm{\odot,init}}$ of the calibrated solar model given by
\citet{Paxton11}, which are 0.2744 and 0.0191 (different
from the present-day abundances of 0.243 and 0.0170), 
and hence the ratio ${\Delta Y}/{\Delta Z}$ is 1.33.
The MESA $\rho$-T tables are based on the 2005
update of OPAL EOS tables \citep{EOS} and opal opacity for the
solar composition of \citet{GS98} supplemented by the
low-temperature opacity \citep{op05} are used. The mixing
length theory of convection is implemented, and $\alpha \equiv
l/H_{\rm{p}}$ is the mixing-length parameter for modulating
convection. Convective overshooting is set as described by
\citet[Section 5.2]{Paxton11} and the overshooting mixing diffusion
coefficient was
\begin{equation}
{D_{\rm{OV}}} = {D_{\rm{conv,0}}}\exp ( - \frac{{2z}}{{f{H
_{p}}}}),
\end{equation}
where $D_{\rm{conv,0}}$ is the mixing
length theory derived diffusion coefficient at a
user-defined location near the Schwarzschild boundary, $z$ is the
distance in the radiation layer away from the location, and $f$ is a
free parameter to change the overshooting scale. 
The photosphere tables is used as the set of boundary
conditions for model atmosphere.

\subsection{Stellar Oscillation Model}
The Aarhus Adiabatic Oscillation Package (ADIPLS) is a simple and
efficient tool for the computation of adiabatic oscillation
frequencies and eigenfunctions for general stellar models
\citep{JCD08, JCD11}. It was used for computing all theoretical
seismic modes in this work. Input parameters for ADIPLS follow the suite
for red giants included in the package. It should note that
structural models generated by MESA (Format FGONG,
{www.astro.up.pt/corot/ntools/docs/Corot\_ESTA\_Files.pdf}) are
redistributed for calculating seismic mixed modes. The number of
structural shells ($N_{\rm{Grid}}$) after redistribution could slightly change the results of frequencies.
We tested some fitting models with $N_{\rm{Grid}}$ from $\sim$2,000 to $\sim$20,000. And computed frequencies become stable when $N_{\rm{Grid}}$ is greater than $\sim$ 6,000.
For all the calculations with ADIPLS, we finally redistributed structural models into
9601 shells. 

\subsection{The Surface Correction}
We used the combined expression and the
method described by \citet{ball14}
for correcting the surface term. 
Based on the discussion of potential asymptotic forms
for frequency shifts \citep{Gough90}, 
the correction formula is a combination of inverse 
and cubic terms:    
\begin{equation}
\delta \nu  = ({a_{-1}}{(\nu /{\nu _{ac}})^{-1}} + {a_3}{(\nu
/{\nu _{ac}})^3})/I,
\end{equation}
where ${a_{-1}}$
and ${a_3}$ are coefficients adjusted to obtain the best frequency 
correction($\delta \nu$). The method to determine these two
coefficients for given set of observed and model frequencies was described
by \citet{ball14}. 
According to the frequency offsets found on the Sun
and other well studied solar-like stars \citep{ball14,kje08},
$\delta \nu$ increases with frequency. In Eq. 6, the surface term 
is also modulated by the normalised mode inertia, $I$. The
description of $I$ can be found in \citet[Eq. (3.140)]{Aerts2010}. 
To avoid confusion, we note that the output results 
from ADIPLS (ADIPLS notes, Eq. 4.3a and 4.3b) in the current 
version is $I/4\pi$.
${\nu _{\rm{ac}}}$ in Eq. 6 is the acoustic cut-off
frequency, which is derived from the scaling relation \citep{Brown91}
\begin{equation}
\frac{\nu _{\rm{ac}}}{\nu _{\rm{ac, \odot}}} \approx \frac{g}{{{g_
\odot}}}\left(\frac{{{T_{\rm{eff}}}}}{{{T_{\rm{eff, \odot }}}}}\right)^{-1/2}.
\end{equation}
Here we take $\nu _{\rm{ac, \odot}}$ = 5000 $\mu$Hz from \citet{ball14}.
Solar references of effective temperature and surface gravity are
$\log g$$_{\odot}= 4.44$ and $T_{\rm{eff},\odot}$ = 5777 K \citep{Cox2000}.

\section{Theoretical Computations and Results}
\subsection{Grid Computation}
The free input parameters for the grid computation
include mass ($M$), initial abundance of heavy elements
($Z_{\rm{init}}$) converted from metallicity ([Fe/H]), initial
abundances of hydrogen and helium ($X_{\rm{init}}$ and
$Y_{\rm{init}}$) which are computed by Eq. 2-4 for a given $Z_{\rm{init}}$,
the mixing-length parameter ($\alpha$), and the overshooting parameter ($f_{\rm{ov}}$). 
We used the masses and their 1-$\sigma$ uncertainties given by \citet{PG16}
as the range of input $M$. 
The input range of [Fe/H] was estimated by both individual 
measurements of ARCES and APOGEE. The lower limit was calculated
with the lower observed [Fe/H] minus its 1-$\sigma$ uncertainty, and the 
upper limit was got from the higher observed [Fe/H] plus its 1-$\sigma$ uncertainty.
For instants, the metallicities of KIC 9970396 given by ARCES and APOGEE
are -0.23$\pm$0.03 and -0.18$\pm$0.07, and hence the input range of
[Fe/H] for this star is from -0.26 to -0.11.
The grid of input $M$ and [Fe/H] were spaced by 0.01$M_{\odot}$ and 0.05 dex, respectively.
We tested the mixing-length parameter ($\alpha$) in a wide range 
around the solar value \citep[$\alpha_{\odot}$ = 1.92]{Paxton11}. 
The grid of $\alpha$ is from 1.72 to 2.52 with a step of 0.1. 
The overshooting parameter $f_{\rm{ov}}$ was either
0.008, 0.012 ,0.016 or 0.020, where the upper limit for this parameter
(0.020) was estimated by \citet{Magic10}.
We used the MESA astero extension's 'grid search' 
function to generate the grid.
In this way, MESA saves structural models that fit observed constraints 
for further analysis. We used four global observables 
of each star as the constraints of the `grid search', namely, 
the average effective temperature and metallicity
from spectra, as well as the surface gravity and radius 
from the binary studies. 
We set a cut-off value of total $\chi^{2}$ as 8.0 for 
selecting MESA models. 
(For a $\chi^{2}$ distribution with 4 degrees of freedom, 
the probability for $\chi^{2}$ smaller than 8 is 0.91.)
A summary of the grid computation with MESA can be found in Table~2. 
ADIPLS was then implemented for calculating radial 
($l$ = 0) and non-radial oscillation modes ($l$ = 1 and 2).

\begin{table*}
 \centering
  \caption{Input Parameters and Observed Constraints for the Grid Computation}
  \begin{tabular}{ccccccccc}
  \hline
KIC & \multicolumn{4}{c}{Grid ranges and Spacing}& \multicolumn{4}{c}{Observed Constraints}\\
\\
    & M/$\delta$M  & [Fe/H]/$\delta$ [Fe/H]  & $\alpha$/$\delta \alpha$ & $f_{\rm{ov}}$/$\delta f_{\rm{ov}}$ & $T_{\rm{eff}}$ &[Fe/H] &$\log g$ & $R$\\
    & [M$_{\odot}$] &[dex] & & &[K]&[dex]&[dex]& [$R_{\odot}$]  \\
  \hline
  4663623 & 1.30 - 1.40/0.01&-0.19 - +0.21/0.05& 1.72 - 2.52/0.10&0.008 - 0.020/0.004&4808(100)&0.01(20)&2.57(3)&9.8(3)\\
  5786154 &1.00 - 1.12/0.01&-0.16 - +0.04/0.05&1.72 - 2.52/0.10 &0.008 - 0.020/0.004&4747(100)&-0.06(06)&2.35(2)&11.4(2)\\
  7037405 &1.21 - 1.29/0.01&-0.44 - -0.04/0.05&1.72 - 2.52/0.10 &0.008 - 0.020/0.004&4529(110)&-0.24(20)&2.24(1)&14.1(2)\\
  8410637 &1.53 - 1.59/0.01&-0.05 - +0.20/0.05&1.72 - 2.52/0.10&0.008 - 0.020/0.004&4750(150)&0.12(17)&2.57(1)&10.7(2)\\
  9540226 &1.28 - 1.38/0.01&-0.38 - -0.08/0.05&1.72 - 2.52/0.10 &0.008 - 0.020/0.004&4677(110)&-0.25(17)&2.349(8)&12.8(1)\\
  9970396 &1.11 - 1.17/0.01&-0.26 - -0.11/0.05&1.72 - 2.52/0.10 &0.008 - 0.020/0.004&4853(150)&-0.20(10)&2.69(2)&8.0(2)\\
  \hline
\end{tabular}
\end{table*}

\subsection{Reordering of Theoretical Mixed Modes}

\begin{figure}
\includegraphics[width=\columnwidth]{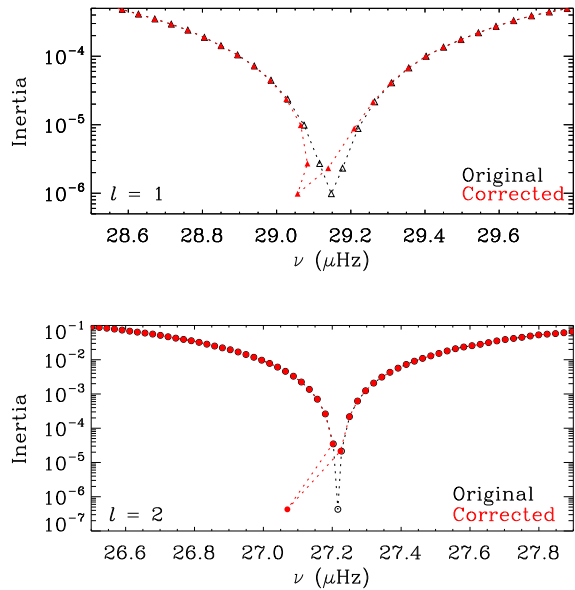}
 \caption{Reordering of mixed modes after surface correction.
          The upper panel includes the mixed modes for $l$ = 1 of
          the best fitting model of KIC 7037405, where black open
          and red filled triangles indicate model frequencies 
          before and after the surface correction.
          The bottom panel is same as the upper one but for $l$ =2.
          The surface correction coefficients for the model are $a_{-1}$ =
          391.6 $\times$ $10^{-9}\mu$Hz and $a_{3}$ = -28.9 $\times$ $10^{-7}\mu$Hz.
         } \label{fig:sc_disorder}
\end{figure}

Before discussing the subsequent fitting progress,  
we address one problem that arises with the current way of applying
the surface correction in evolved stars: theoretical mixed modes could 
lose their original order after their surface terms are corrected.
We took the model of KIC 7037405 as an example.
Fig. \ref{fig:sc_disorder} shows the mixed modes
for $l$ = 1 and 2 in a single order of this model.
Theoretical mode inertia are plotted against 
frequencies before and after the surface correction. 
Due to the quick change of inertia, p-dominated modes
(with the lower inertia) get much larger corrections than their
g-dominated neighbours. And the modes
obtain a new order after the correction. 
Similar reordering appears in theoretical $l$ = 1 mixed modes of
KIC 5786154, KIC 7037405, and KIC 9940226 and in the $l=2$ modes of all six targets. 
The re-ordering happens because the surface corrections for the mixed 
modes differ by more than the separation between consecutive modes.
The g-like modes are less changed by the surface properties
than p-like modes due to their large mode inertia. 
And hence we do not expect a strong surface term for them.    
However, one thing worth to note is that theoretical mixed modes
are solved as the coupling results of p and g modes. 
If the surface term of the p-mode is not being corrected before 
calculating the coupling, the mixed modes will also be affected. 

\begin{figure}
\includegraphics[width=\columnwidth]{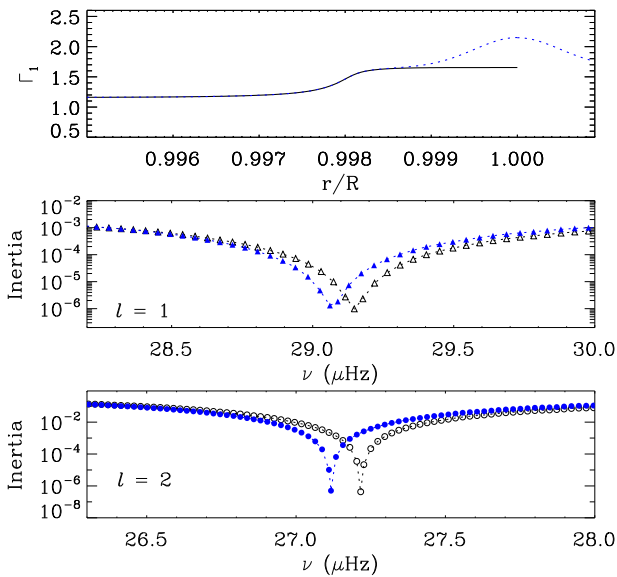}
 \caption{The change of frequency ranges where
          the p-g coupling happens before and after
          modifying the structures of the near-surface 
          layers. Top: the distribution of 
          $\Gamma _{1}$ at the near-surface layers 
          before (black solid) and after (blue dots) the modification. 
          The fractional increase of $\Gamma _{1}$ with 
          the depth (r/R) follows a Gaussian function.
          The centre, width, and amplitude of this Gaussian
          profile are 1.0R, 0.0005R, and 0.3, respectively.
          Middle: mixed modes for $l$ = 1 of the models 
          before (black open triangles) and after (blue filled triangles) 
          the modification. 
          Bottom: mixed modes for $l$ = 2 before (black open circles) 
          and after (blue filled circles) 
          the modification.
         } \label{fig:modify_fgong}
\end{figure}

A proper way to test this issue could be correcting the acoustic wave
before calculating the mixed modes. However, no option is available in 
the seismic code to do this job. Instead, we changed 
the structure in the near-surface layers of the input model of ADIPLS. 
This only alters the acoustic wave, 
but not the gravity modes because they travel 
inside the core. Modifying the adiabatic exponent 
($\Gamma _{1}$ = ($d$ln$p$/$d$ln$\rho$)$_{\rm{ad}}$) in the
near-surface layers is a simple way to modulate the acoustic resonance.
However, it should be note that the structural model after this modification 
is no longer self-consistent. 
For the case of the Sun, the surface term starts
from 1Mm below its surface where $T = T_{\rm{eff}}$. 
The depth of this region is roughly 0.15\% of the total solar radius.
Hence we only changed $\Gamma _{1}$
in layers that are 0.15\% of the total radius below the surface. 
As shown in the top graph of Figure \ref{fig:modify_fgong},
we gradually increased $\Gamma _{1}$ in the region 
from 0.9985 $R$ to the surface. 
The fractional increase at given depth followed a Gaussian function 
(blue dotted line).
The centre of the Gaussian profile is right on the top layer. 
And its amplitude and width are 0.3 and 0.0005$R$, respectively.
The mixed modes for $l$ = 1 and 2 calculated with 
the models before and after modifying the surface-corrected $\Gamma_{1}$
are also given in the middle and bottom panels. 
It can be found that the change in surface layers affects the most p-like modes. 
Shifted p modes then change the frequency ranges where p-g coupling happens. 
Thus, the surface term also affects the mixed modes in an indirectly way.

Modifying the stellar structure seems to be
a way to repair the indirect effects of the 
surface term on mixed modes. However, 
it comes with difficulties in rebuilding the equilibrium
of the model. 
We hence still adopted the correcting formula (Eq. 6) 
to correct model frequencies. 
To avoid the influence when using it in the following fitting procedures,
we checked the model frequencies for all six stars.
Hence only the most-p-like modes are considered for these cases.
The $l$ = 1 mixed modes of the other three stars 
(KIC4663623, KIC8410637, and KIC9970396)
have large enough period spacing to overcome the 
surface effect without mode order swapping and were used for the following studies.
Based on the models of six red giants, 
$\nu_{\rm{max}}$ greater than $\sim$ 40 $\mu$Hz tends to be 
a safe cut for stars whose mixed modes for $l$ = 1 can be adopted.  
We will mention the identification of observed individual mixed modes 
for these stars below.    

\subsection{Identification of Individual Mixed Modes}
Peak-bagging of individual mixed modes for red giants is 
complicated by rotational splitting, 
granulation background, and random noise. 
On the other side, the
theoretical models provide precise oscillation frequencies of 
mixed modes for every
degree and order, which can be a guide
to peak-bagging. 

\begin{figure*}
\includegraphics{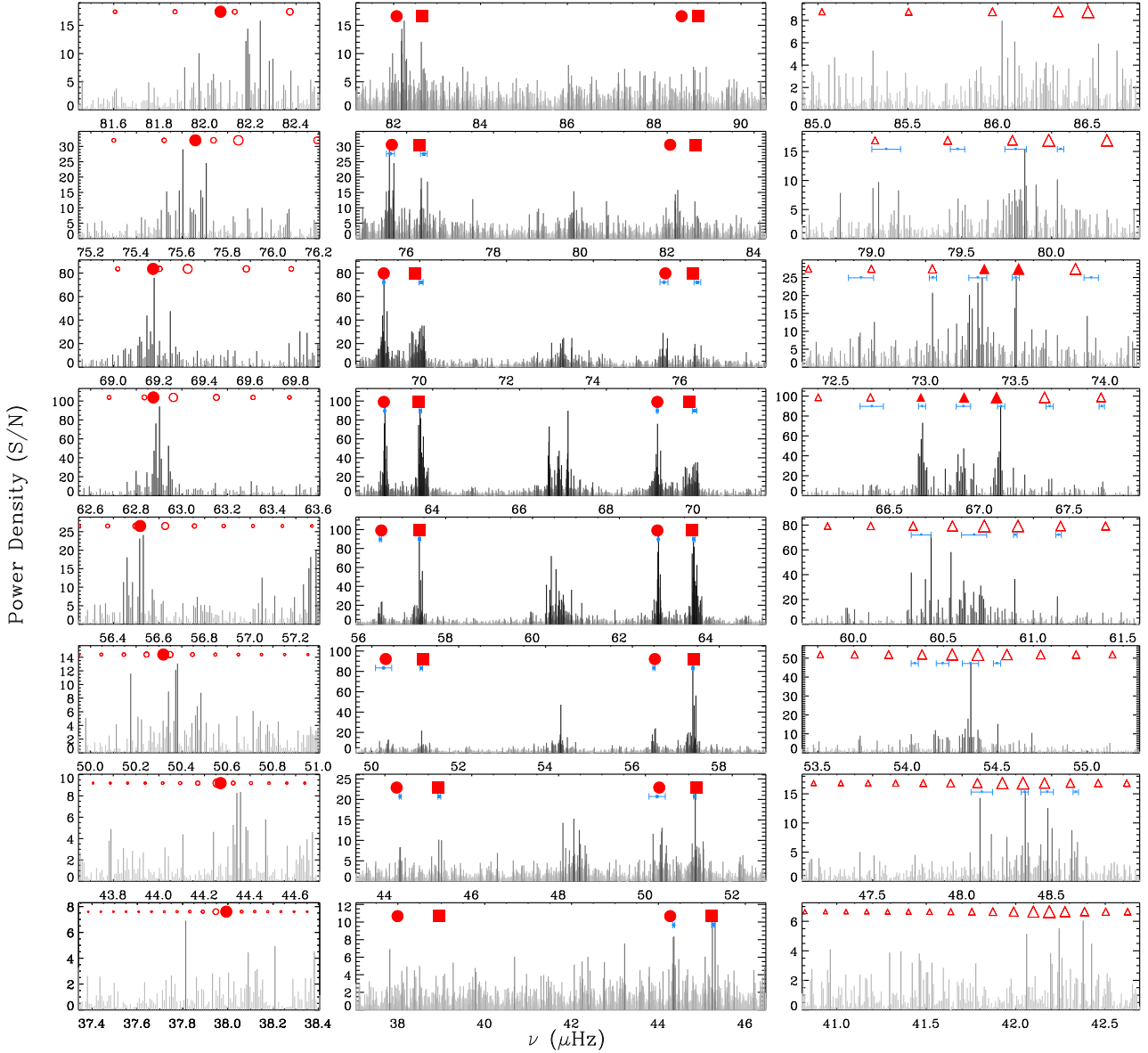}
 \caption{The peak-bagging process of individual mixed modes for KIC 9970396.
The whole power spectrum is separated into eight radial-mode orders
as shown in the middle. Close inspections of $l$ = 2 and 1 modes are plotted on the left and right sides. 
The colour code was set as same as that in Figure 1, 
indicating the $\mathcal{P}_{\rm{signal}}$
of each frequency bin.
Red Symbols plotted on the top are theoretical
frequencies of the best fitting model. Squares, circles and
triangles in the middle represent the p and most-p-like modes for $l$ = 0, 1 and 2.
Filled symbols are the modes for picking the fitting models, and 
open symbols are those for guiding the identification of other mixed modes.
Circles and triangles on the left and right indicate all theoretical mixed modes in each frequency bin. And their symbol size is scaled 
with $1/I^{2}$ ($I$ is mode inertia) by reference to that of the
most p-like mode in each degree and order.
Larger size indicates the mode to be more p-like and less in inertia.     
Small blue symbols represent identified observed frequencies.
} \label{fig:ps}
\end{figure*}

We firstly found theoretical models which
fit the observed modes for $l$ = 0 and 2 (the model
with the highest 10\% likelihood).
These models then guided us to identify $l$ = 1 modes.  
One example, as given in Fig. \ref{fig:ps}, illustrates
the peak-bagging process of mixed modes for KIC9970396.
The model was firstly constrained by the modes
for $l$ = 0 and 2 given in the middle panels,
as well as a few high-amplitude
mixed modes for $l$ = 1 (filled symbols), 
such as the three peaks from 66.7 to 67.2 $\mu$Hz. 
All model frequencies were then plotted on the power spectrum 
(open triangles),
suggesting more potential modes like the one at 66.4 $\mu$Hz, and
two at 67.4 and 67.7 $\mu$Hz. 
It should be noted that the power of these modes is still 
significant compared with the background noise level.
The method we use for getting the frequencies and uncertainties
of individual mixed modes are as follow. 
For a potential mode based on the visual inspection, 
we flagged the observed peaks with 
$\mathcal{P}_{\rm{signal}}$ > 0.5 around the theoretical predictions. 
The median value of these peaks was adopted as the frequency
of this mode. And the difference between the median value and 
the highest/lowest frequency was measured as the uncertainty.
The blue error bars in Figure 1 represent all of the
identified modes and their uncertainties.
We also tried to identify the g-dominated modes for
$l$ = 2 (on the left side). However, no significant
mode has been found because of their large inertia. 
All the identified modes of KIC 9970396 are listed in Table~3, 
The power spectra and tables of 
all six stars can be found in the Appendix.

\begin{table}
 \centering
  \caption{Identified Oscillation Frequencies for Star KIC 9970396}
  \begin{tabular}{cccc}
  \hline
$l$ & $\nu$ [$\mu$Hz] & $\sigma$ [$\mu$Hz] \\
  \hline
       2 &    44.351 &   0.021  \\  
      0 &    45.255 &   0.032  \\  
      2 &    50.270 &   0.187  \\  
      0 &    51.138 &   0.029  \\  
      2 &    56.496 &   0.023  \\  
      0 &    57.396 &   0.015  \\  
      2 &    62.895 &   0.012  \\  
      0 &    63.717 &   0.023  \\  
      2 &    69.173 &   0.017  \\  
      0 &    70.031 &   0.050  \\  
      2 &    75.626 &   0.091  \\  
      0 &    76.395 &   0.077  \\
     1  &    48.110 &   0.060  \\
     1  &    48.350 &   0.020  \\
     1  &    48.475 &   0.034  \\       
     1  &    48.635 &   0.016  \\
     1  &    54.040 &   0.018  \\               
     1  &    54.195 &   0.034  \\  
     1  &    54.350 &   0.016  \\
     1  &    54.498 &   0.020  \\
     1  &    54.703 &   0.029  \\
     1  &    60.420 &   0.110  \\  
     1  &    60.650 &   0.052  \\  
     1  &    60.899 &   0.012  \\
     1  &    61.140 &   0.015  \\
     1  &    66.400 &   0.065  \\     
     1  &    66.680 &   0.020  \\
     1  &   66.910  &  0.040   \\
     1  &    67.120 &   0.020  \\
     1  &    67.390 &   0.023  \\    
     1  &    67.680 &   0.015  \\
     1  &    72.640 &   0.070  \\     
     1  &    73.040 &   0.022  \\           
     1  &    73.290 &   0.051  \\
     1  &    73.501 &   0.019  \\     
     1  &    73.921 &   0.041  \\  
     1  &    79.100 &   0.100  \\
     1  &    79.477 &   0.040  \\
     1  &    79.850 &   0.050  \\
     1  &    80.032 &   0.032  \\
                            
  \hline
\end{tabular}
\end{table}

\subsection{Calibrating the Model Parameters}

We used a Bayesian method to estimate the mixing-length parameter ($\alpha$) and
the two coefficients for surface correction ($a_{-1}$ and $a_{3}$). 
The agreement between models and observations was first examined
by a likelihood function. The observed data for these stars can be divided
into three parts. The mass, radius, and atmospheres provide global features;
p and the most p-like modes mainly relate to the structure of the envelope; and
p-g mixed modes probe the characteristics of the core. Thus, the likelihood 
($\mathcal{L}$) for every model comprises these three parts, namely, $\mathcal{L}_{\rm{glob}}$, 
$\mathcal{L}_{\rm{env}}$ and $\mathcal{L}_{\rm{core}}$.

\begin{figure}
\includegraphics[width=\columnwidth]{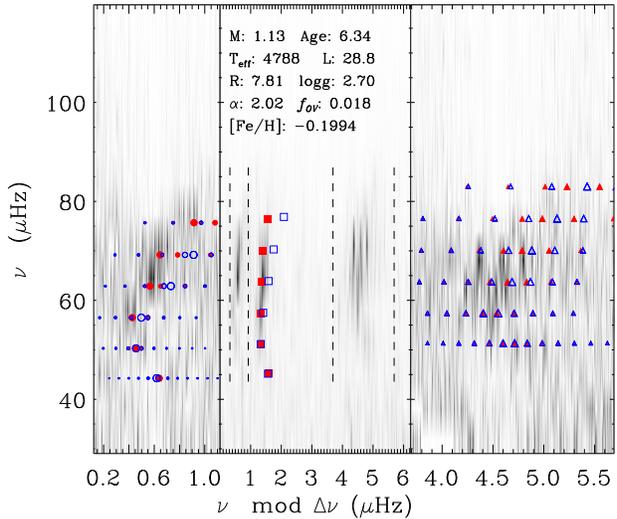}
 \caption{Observed and theoretical \'echelle diagram of KIC 9970396. 
 Grey-scale is the observed power density. Blue and red symbols
 indicate model frequencies before and after surface correction. 
 The middle panel includes the whole \'echelle diagram, 
 where red squares are radial modes. 
 Left and right panels zoom in on $l$ = 2 and 1
modes. Theoretical mixed modes are plotted in different sizes 
scaled by $1/I^2$ ($I$ is the normalized mode inertia) by reference
to the most p-like mode at same degree and order. The largest symbol indicates the
most p-like mode for $l$ = 1 and 2 in each order.} \label{fig:ed}
\end{figure}

The likelihood of global features was calculated using the masses, radii, 
and surface gravities from dynamical modelling, as well as effective temperatures
($T_{\rm{eff}}$) and metallicities ([Fe/H]) from spectra. 
It is described as
\begin{equation}
\mathcal{L} _{\rm{glob}}{\rm{ = }}\mathrm{exp}\left(-\frac{{{1}}}{2n}\sum\limits_{i = 1}^n
{\frac{{{{({x_i} - {u_i})}^2}}}{{\sigma _i^2}}}\right),
\end{equation}
where ${x_i}$ and ${u_i}$ indicate theoretical and
observed parameters, and ${\sigma _i}$ is the uncertainties of observations.
The likelihood for the envelope was estimated by comparing p and 
the most p-like mixed modes from models and observations: 
  \begin{equation}
\mathcal{L} _{\rm{env}}{\rm{ = }}\mathrm{exp}\left(-\frac{{{1}}}{2n}\sum\limits_{i
= 1}^n {\frac{{{{({\nu_{{model},i}} - {\nu_{{obs},i}})}^2}}}{{\sigma
_{{obs},i}^2}}}\right).
\end{equation}
Individual g-dominated mixed modes were identified for KIC 4663623, KIC 8410637, and KIC 9970396, 
(the other three have the problem with the surface correction) and we used them to estimate the likelihood for the core by
\begin{equation}
\mathcal{L}_{\rm{core}}{\rm{ = }}\mathrm{exp}\left(-\frac{{{1}}}{2n}\sum\limits_{i
= 1}^n {\frac{{{{({\nu_{{model},i}} - {\nu_{{obs},i}})}^2}}}{{\sigma
_{{obs},i}^2}}}\right).
\end{equation}
These three parts gave the final likelihood, described as
\begin{equation}
\mathcal{L}{\rm{ = }}\mathcal{L} _{\rm{glob}}\mathcal{L} _{\rm{env}}\mathcal{L} _{\rm{core}}.
\end{equation}
The best-fitting model of KIC 9970396 based on the likelihood test
is given in Fig. \ref{fig:ed}.
It shows that model frequencies fit quite 
well after the surface correction. 

The theoretical models in this work have four independent input parameters, 
namely, mass ($M$), heavy-element abundance ($Z$), the mixing-length parameter ($\alpha$)
and overshooting parameter ($f_{\rm{ov}}$). Along with age ($\tau$) and
two coefficients ($a_{-1}$ and $a_{3}$), these seven variables specify a 
particular model. Thus, we can write the probability of a model as
\begin{equation}
\it{p}(\rm{Model|Data}){\rm{ = }} \it{p}(\rm{Model})\it{p}(\rm{Data|Model}) {\rm{ = }} \rm{prior}\cdot \mathcal{L},
\end{equation}
where Model = \{M, Z, $\tau$, $\alpha$, $f_{\rm{ov}}$, $a_{\rm{-1}}$, $a_{\rm{3}}$\}
and Data = \{$T_{\rm{eff}}$, $\log g$ ,[Fe/H], R, $\nu_{l,n}$\}.
$\mathcal{L}$ is the likelihood obtained above. 
We assumed a flat prior for all model quantities. 
Then the probability of $\alpha$, $a_{-1}$ and $a_{3}$ for a star can be 
calculated by marginalising over all the other model quantities:
\begin{equation}
p(\alpha){\rm{ = }}\int{p(M, Z, \tau, \alpha, f_{\rm{ov}}, a_{-1}, a_{3})}dM dZ d\tau df_{\rm{ov}} da_{-1} da_{3},
\end{equation}
\begin{equation}
p(a_{-1}){\rm{ = }}\int{p(M, Z, \tau, \alpha, f_{\rm{ov}}, a_{-1}, a_{3})}dM dZ d\tau d\alpha df_{\rm{ov}} da_{3}
\end{equation}
and
\begin{equation}
p(a_{3}){\rm{ = }}\int{p(M, Z, \tau, \alpha, f_{\rm{ov}}, a_{-1}, a_{3})}dM dZ d\tau d\alpha df_{\rm{ov}} da_{-1}.
\end{equation}

\begin{figure*}
\includegraphics[scale=0.66]{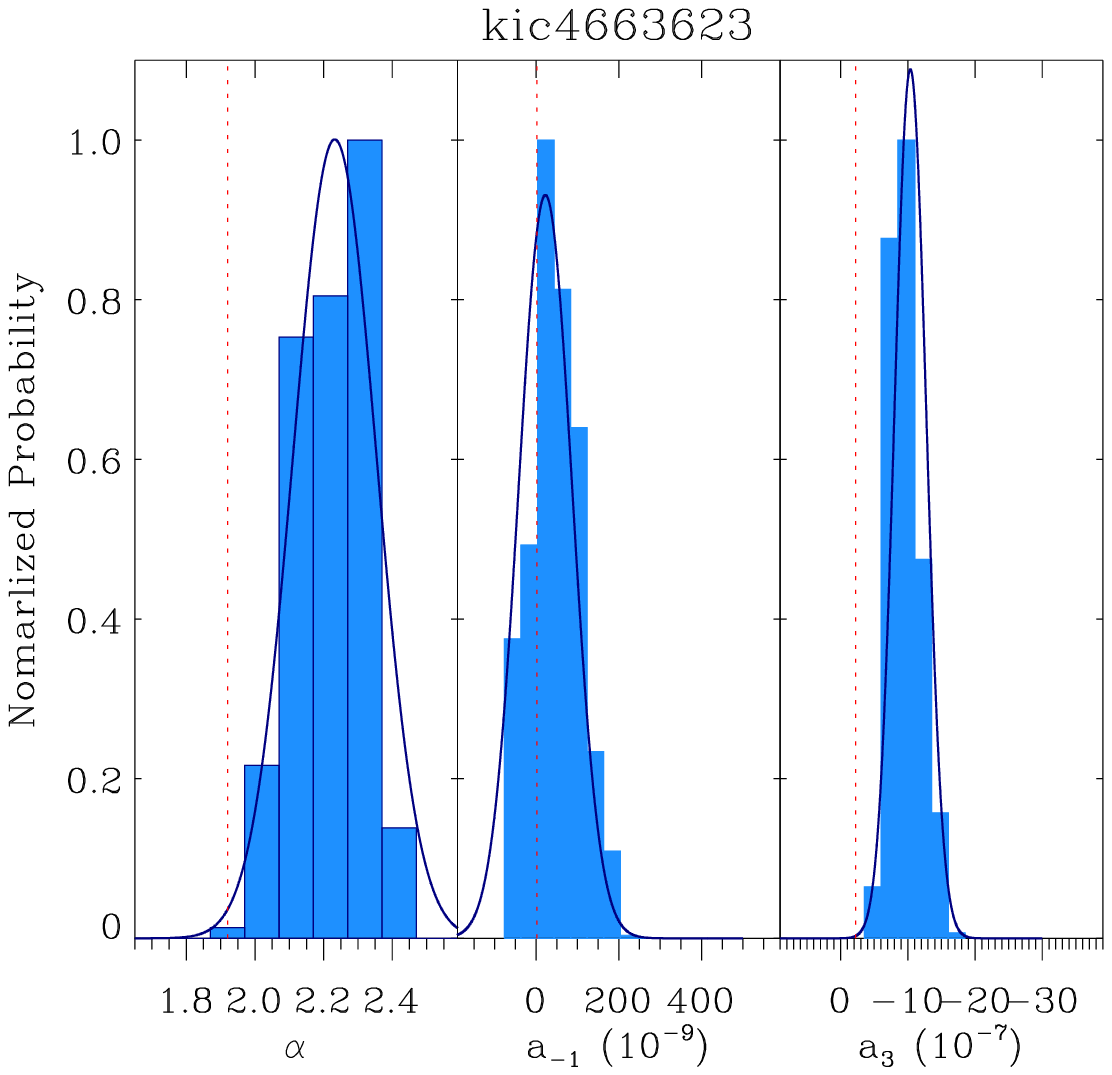}
\includegraphics[scale=0.66]{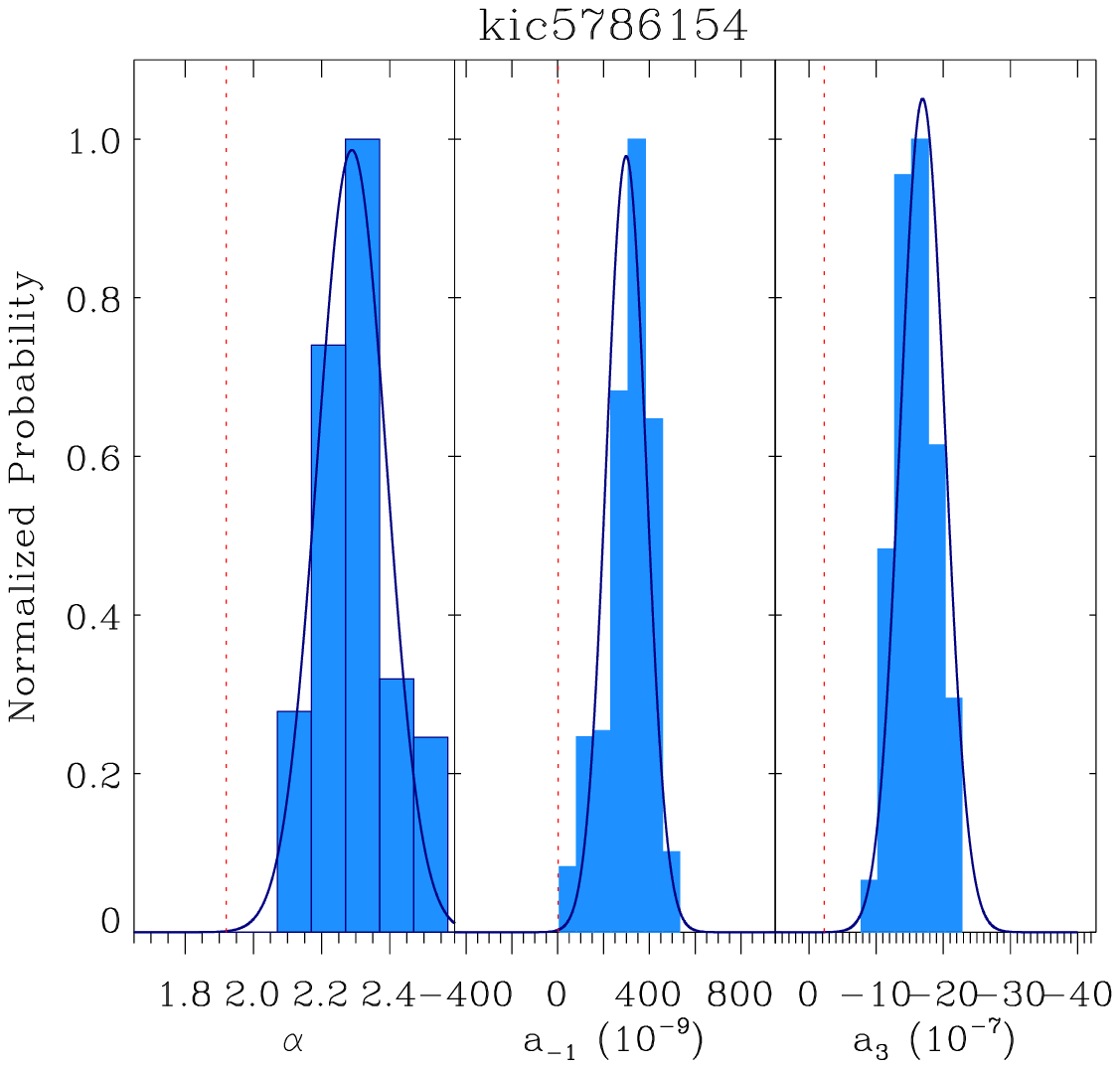}
\includegraphics[scale=0.66]{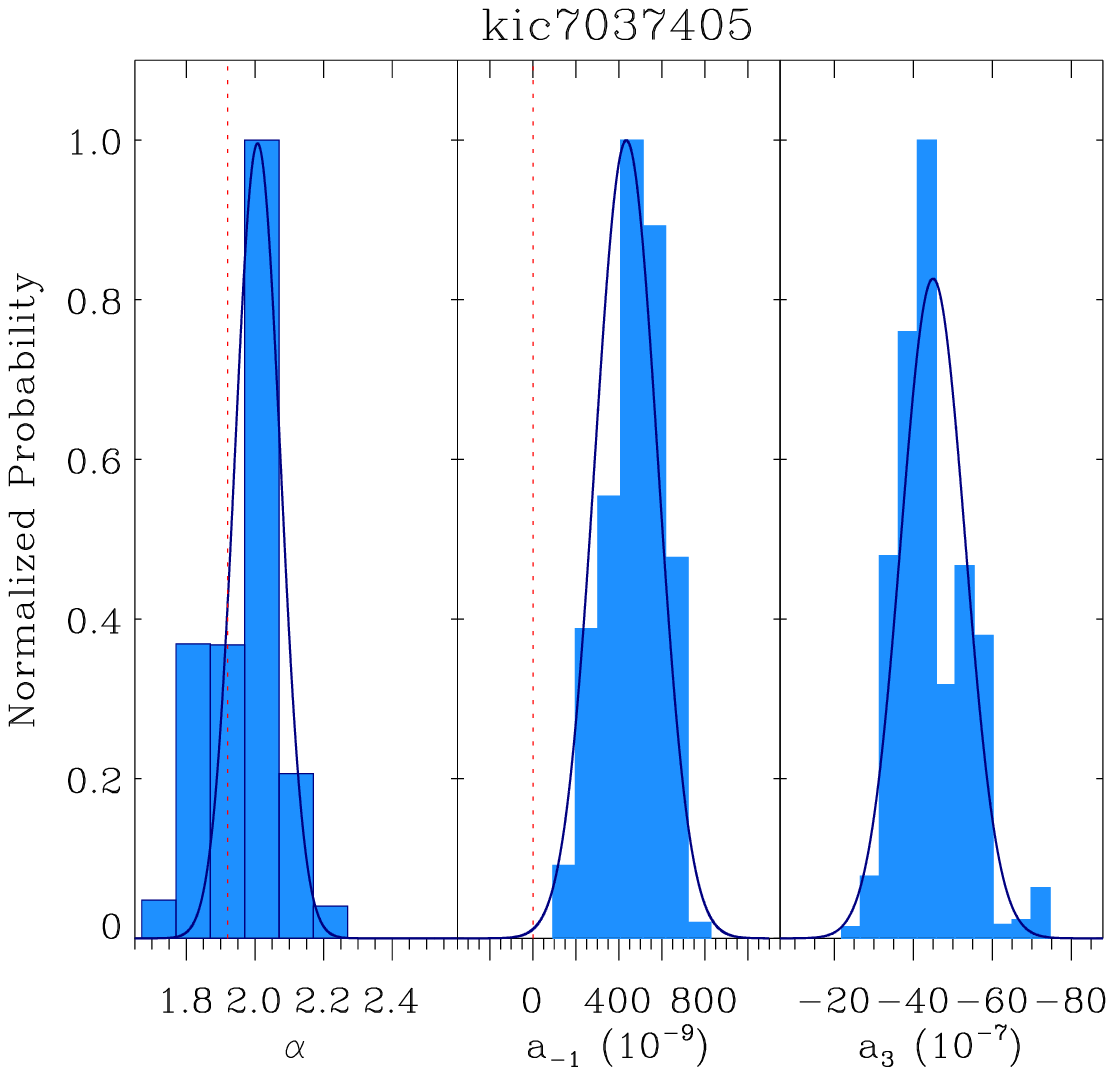}
\includegraphics[scale=0.66]{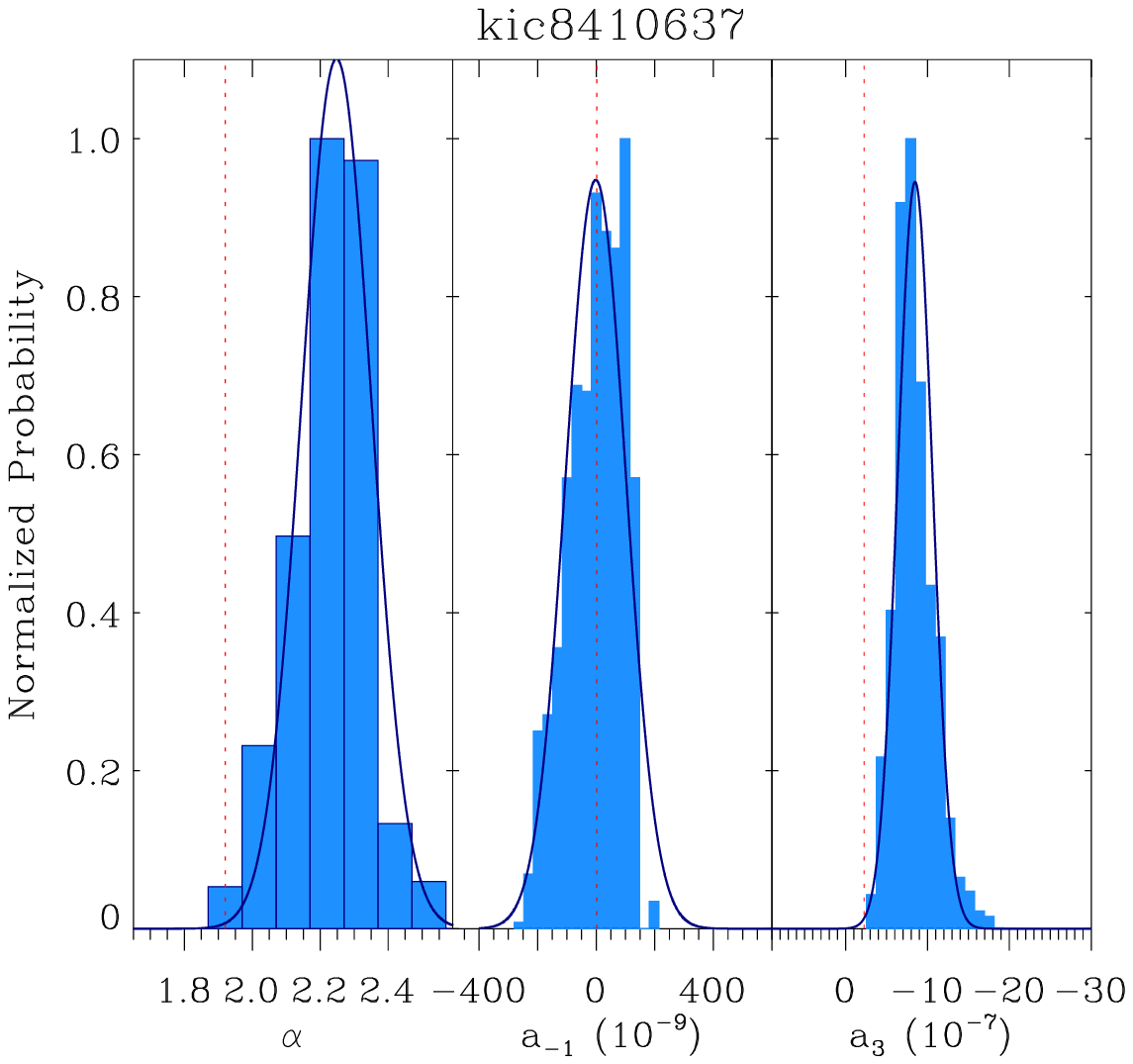}
\includegraphics[scale=0.66]{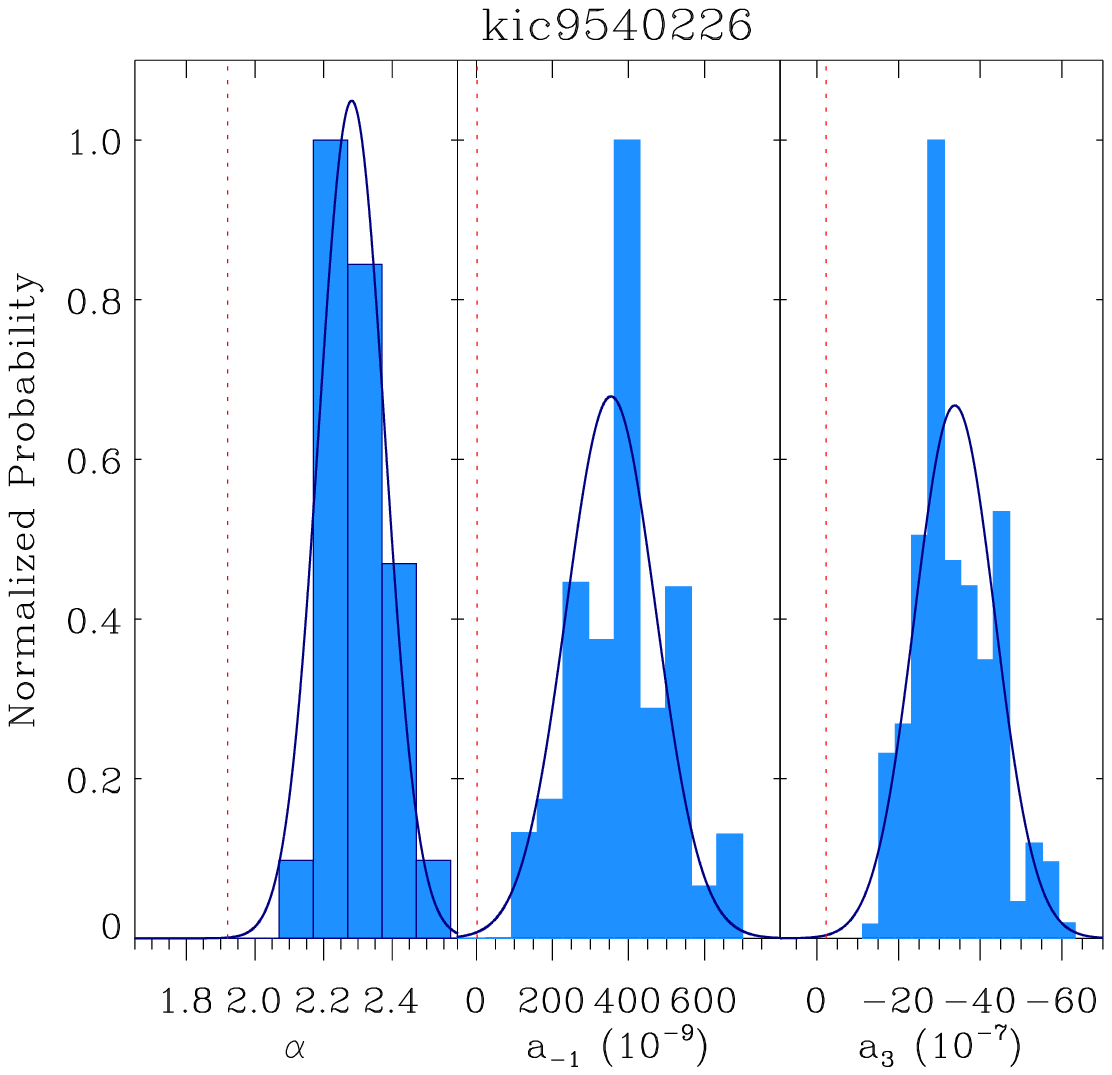}
\includegraphics[scale=0.66]{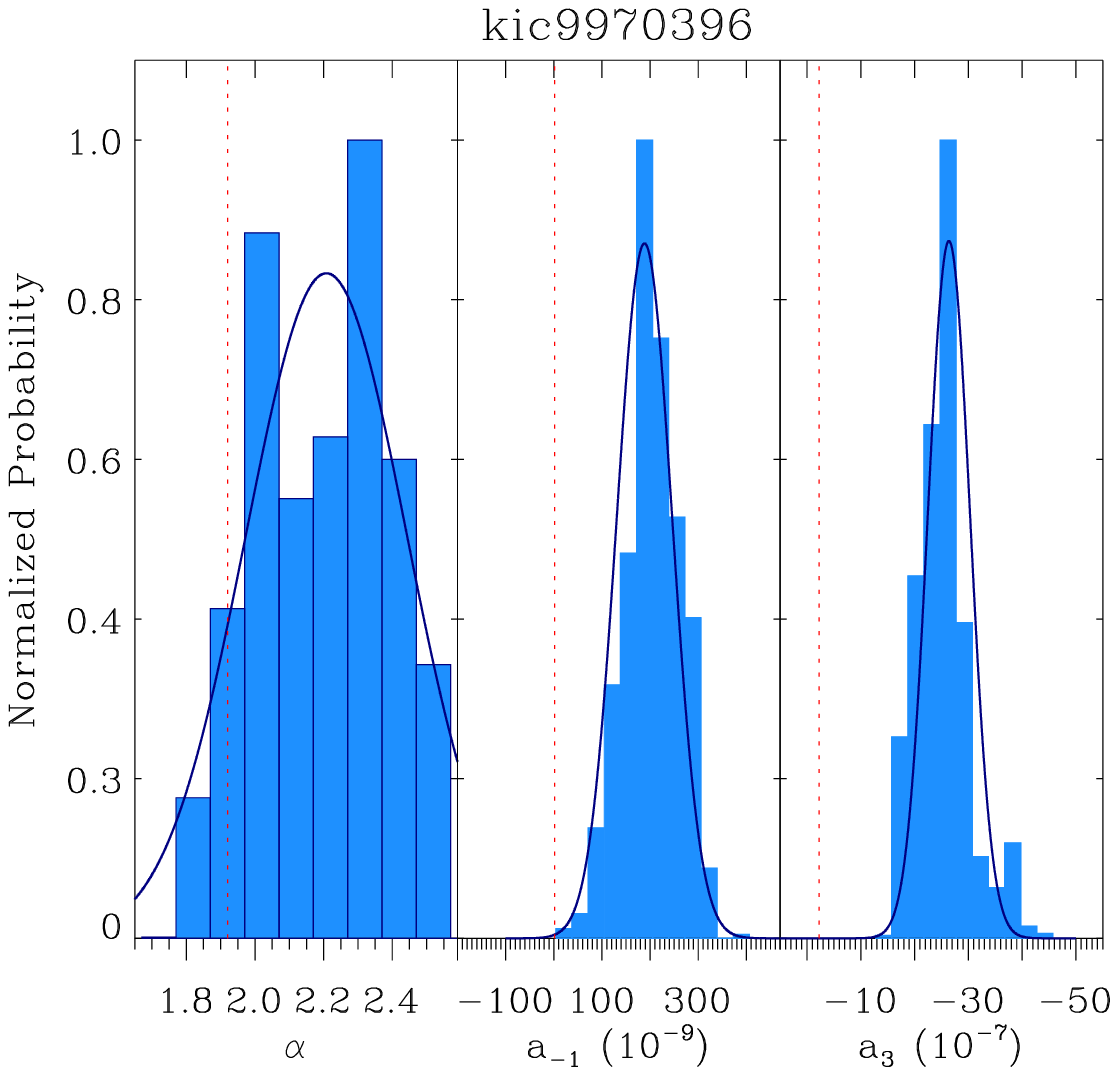}
 \caption{Probability distributions for the mixing-length parameter ($\alpha$) 
 and the two coefficients ($a_{-1}$ and $a_{3}$) for the surface correction. 
 Histograms are normalised proportional to the highest peak.
  $a_{-1}$ and $a_{3}$ are scaled by $10^{-9}$ and $10^{-7}$ $\mu$Hz.
 Solid curves are the Gaussian profiles for fitting the probability distributions.
 Red dotted lines indicate solar values, which are 1.92 for 
$\alpha$ \citep{Paxton11}, $1.73 \times 10^{-9}\mu$Hz for $a_{-1}$ and 
$-2.25 \times 10^{-7}\mu$Hz for $a_{3}$ \citep{ball14}.} 
\label{fig:a-a1}
\end{figure*}

Fig. \ref{fig:a-a1} gives the probability distributions
of $\alpha$, $a_{-1}$, and $a_{3}$ for the six stars. 
The histograms mainly follow Gaussian distributions.
Some profiles have a sudden drop rather than a smooth decrease
at the boundary. The reason is that the MESA astero 'grid search' 
function was set to only save models 
that were in reasonable agreement with the observations ($\chi^{2}<8$).
We used a Gaussian function to fit probability distributions,
and adopted the centre and 1-$\sigma$ deviation of the Gaussian profile as
the central value and the uncertainty.
The results are summarised in Table~4.
 
 \begin{table}
   \caption{Calibrated Mixing-Length Parameter and the Two Coefficients in the Surface-Correction Expression (Eq. 6)
   }
    \resizebox{\columnwidth}{!}{
   \begin{tabular}{cccc}
   \hline
 Star & $\alpha$  & $a_{-1}$ & $a_{3}$  \\
     &           &[$10^{-9}\mu$Hz]&[$10^{-7}\mu$Hz]\\
   \hline
   Sun&1.92&1.73&-2.25\\
   KIC 4663623&2.23$\pm$0.12&22.50$\pm$64.74&-10.4$\pm$2.4\\
   KIC 5786154&2.29$\pm$0.10&298.9$\pm$87.0&-16.9$\pm$3.3\\
   KIC 7037405&2.01$\pm$0.07&435.0$\pm$145.1&-45.0$\pm$8.1\\
   KIC 8410637& 2.25$\pm$0.10&-1.6$\pm$102.7&-8.5$\pm$2.1\\
   KIC 9540226&2.28$\pm$0.10&354.0$\pm$117.7& -33.8$\pm$9.8\\
   KIC 9970396&2.21$\pm$0.23&188.3$\pm$57.5& -26.4 $\pm$3.9\\
   \hline
 \end{tabular}
 }
 \end{table}

\section{Model Parameters}

\subsection{The Mixing-Length Parameter}

The calibrated mixing-length parameters
of the six red giants (Table 4) 
suggest larger $\alpha$ for evolved stars than 
that for the Sun ($\alpha_{\odot}$ = 1.92).
The average mixing-length parameter is 
about 2.20. However, KIC 7037405 converges
at significantly different values (2.00). 
As mentioned, the determination of $\alpha$ depends greatly
on the measurements of the mass (from the binary orbits). 
Modelling a star with increased mass tends to lead to
a smaller mixing-length parameter when other constraints
are kept fixed. Hence systematic 
errors in mass can bias the results. 
KIC 7037405, 9540226, and 9970396 were 
also recently studied by \citet{Brogaard17}. 
They obtained masses
that differ by -2$\sigma$ for KIC 7037405, 
0.96$\sigma$ for KIC 9540226, and 1.25$\sigma$ for KIC 9970396. 
The results suggest that KIC 7037405 could have 
a larger mixing-length parameter than what was obtained (2.00) 
while KIC 9540226 and KIC 9970396 may converge 
at a slightly smaller $\alpha$. 

The average $\alpha$ obtained for our six giants 
is $\sim$1.14 $\pm$ 0.07 times of the calibrated solar value. 
This result is similar to that achieved
by \citet{Tayar17} with the APOGEE-$Kepler$ targets. 
Their grid models for the red giants within the range of metallicity ([Fe/H]) 
from $-0.5$ to $+0.4$ required a $\sim$8\% larger $\alpha$ 
than the Sun to fit the observations. 
\citet{Tayar17} also suggested a linear correlation 
between $\alpha$ and [Fe/H]. However, we did not obtain a clear 
dependence between the two parameters in our stars.  

The realistic simulation of convection is another 
approach to estimate the mixing-length parameter.
\citet{Magic15} calibrated $\alpha$ with the STAGGER
grid, which predicted slightly smaller values (0.90 - 0.95 $\alpha _{\odot}$) 
for red giants ($T_{\rm{eff}}$ $\simeq$ 5,000K and $\log g$ $\simeq$ 2.5) than that for the Sun. An earlier work by \citet{Trampedach11} found the mixing-length parameter to be
0.96 - 1.00 $\alpha _{\odot}$ for the red giants having similar 
stellar parameters with our stars 
(M = 0.8 - 2.5 $M_{\odot}$, $T_{\rm{eff}}$ = 4200 -- 5500K, $\log g$ = 2.4 -- 3.0)
through 3-D simulations \citep{Trampedach14}.
The $\sim$16\% difference of the 1-D stellar model from 
the simulations may indicate an improper modelling for 
the near-surface layers in the 1-D model with the 
current input physics. Firstly, the mixing-length 
parameter of a star varies for different evolutionary 
stages \citep{Magic15,Trampedach14}, however, we 
fixed the mixing-length parameter through the 
stellar evolution. Secondly, the boundary conditions
for red giants can be different from the Sun due to the 
changes in structures. Moreover, the systematic offsets
in observed effective temperatures and metallicities 
could also affect the results.

 \begin{figure*}
 \includegraphics[scale=0.45]{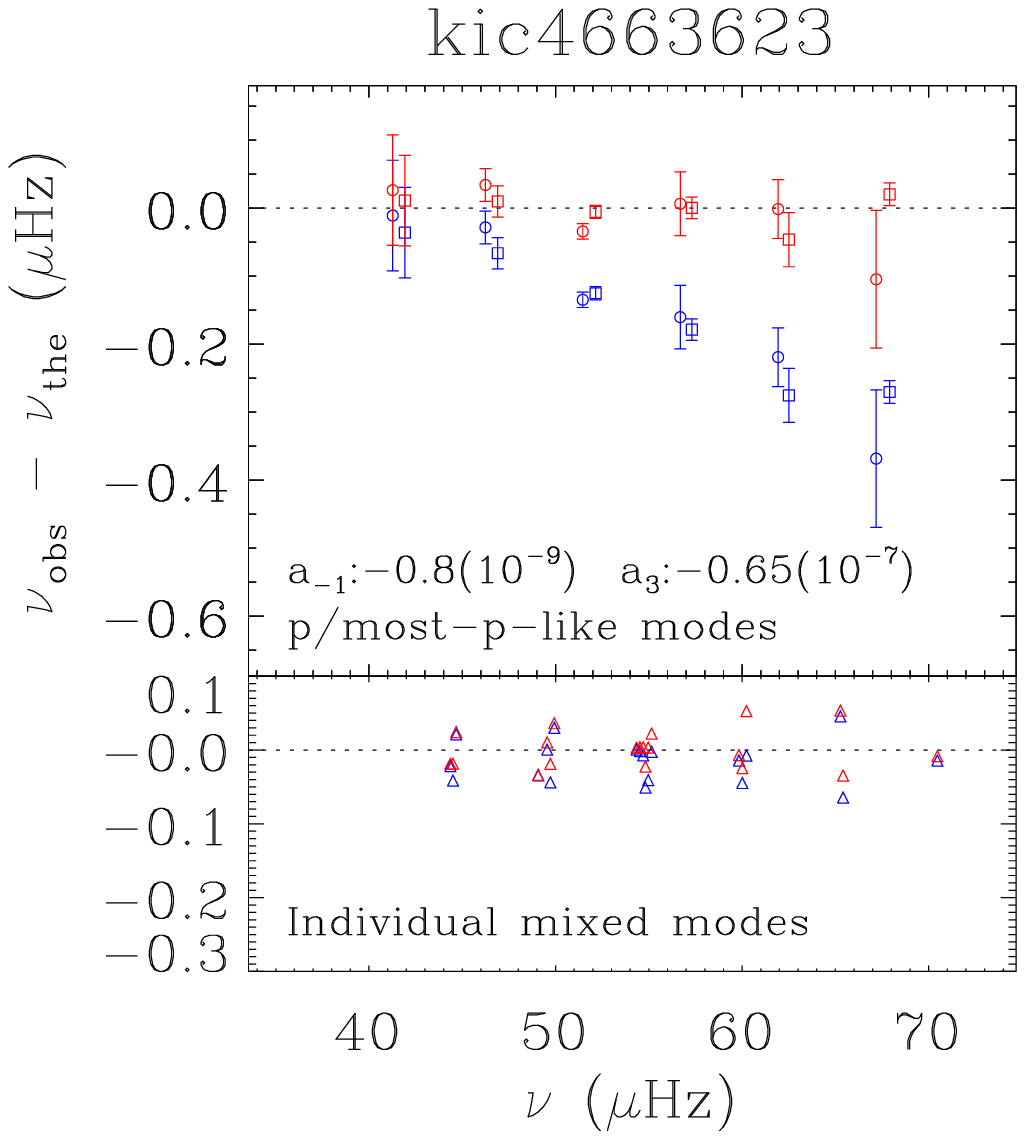}
  \includegraphics[scale=0.45]{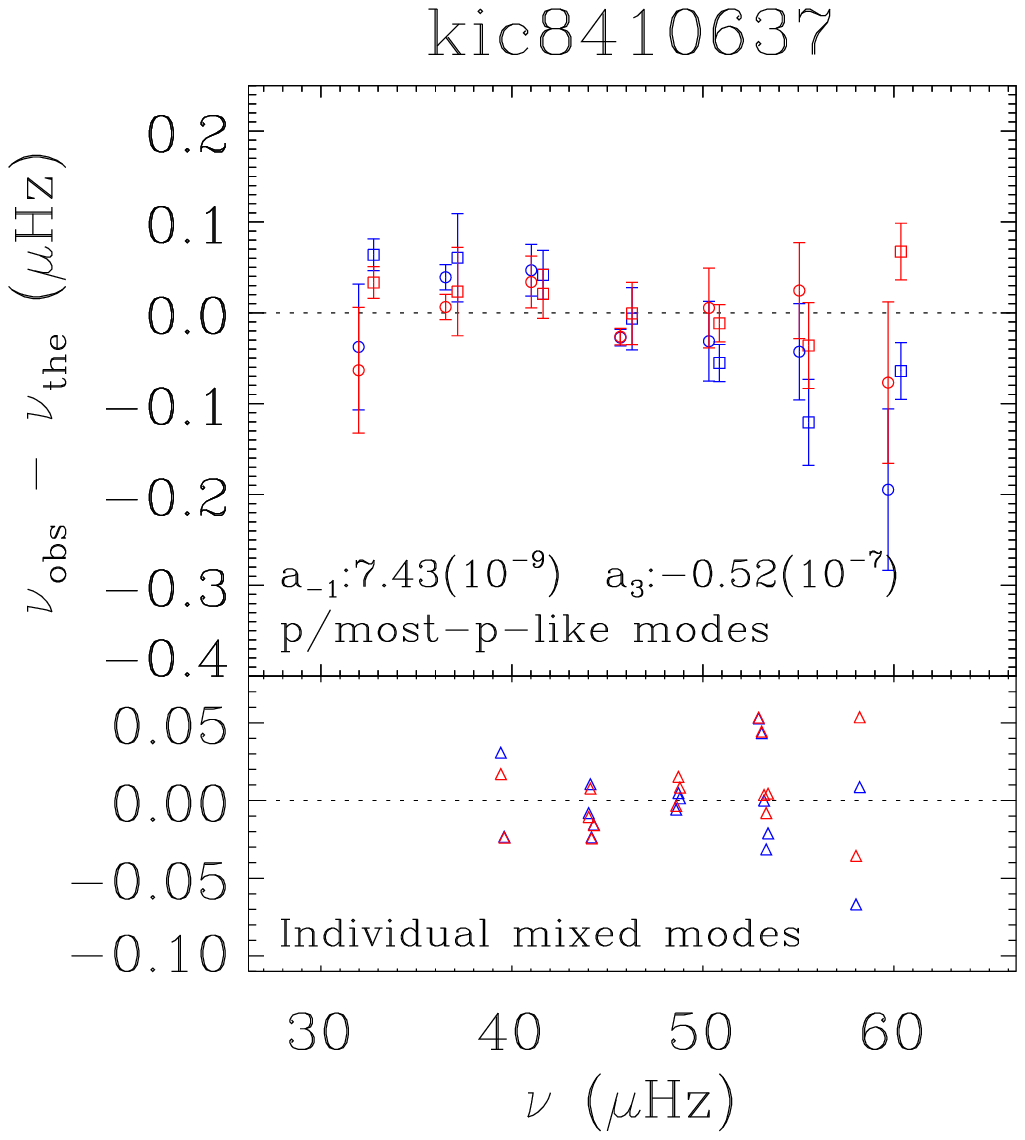}
   \includegraphics[scale=0.45]{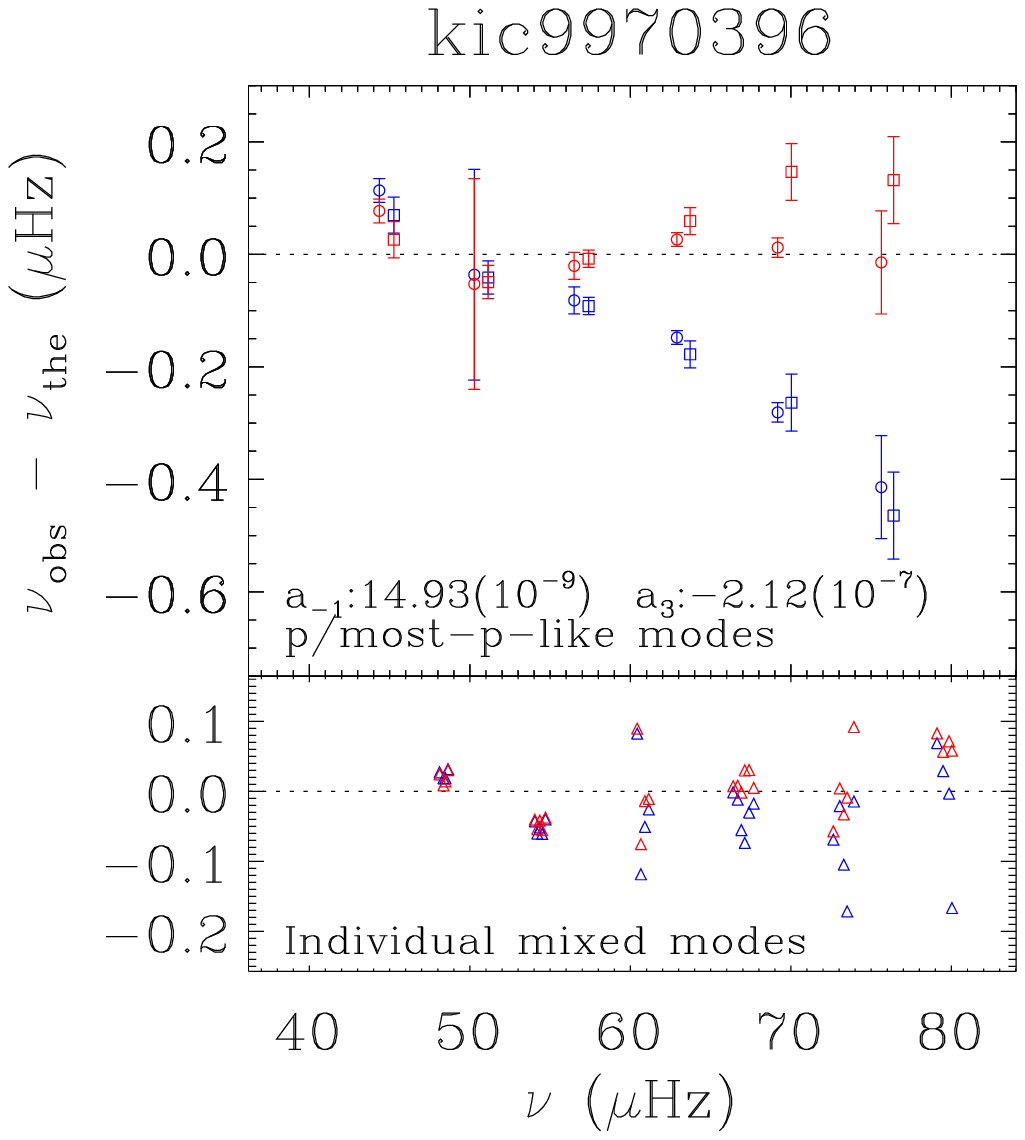}
 \includegraphics[scale=0.45]{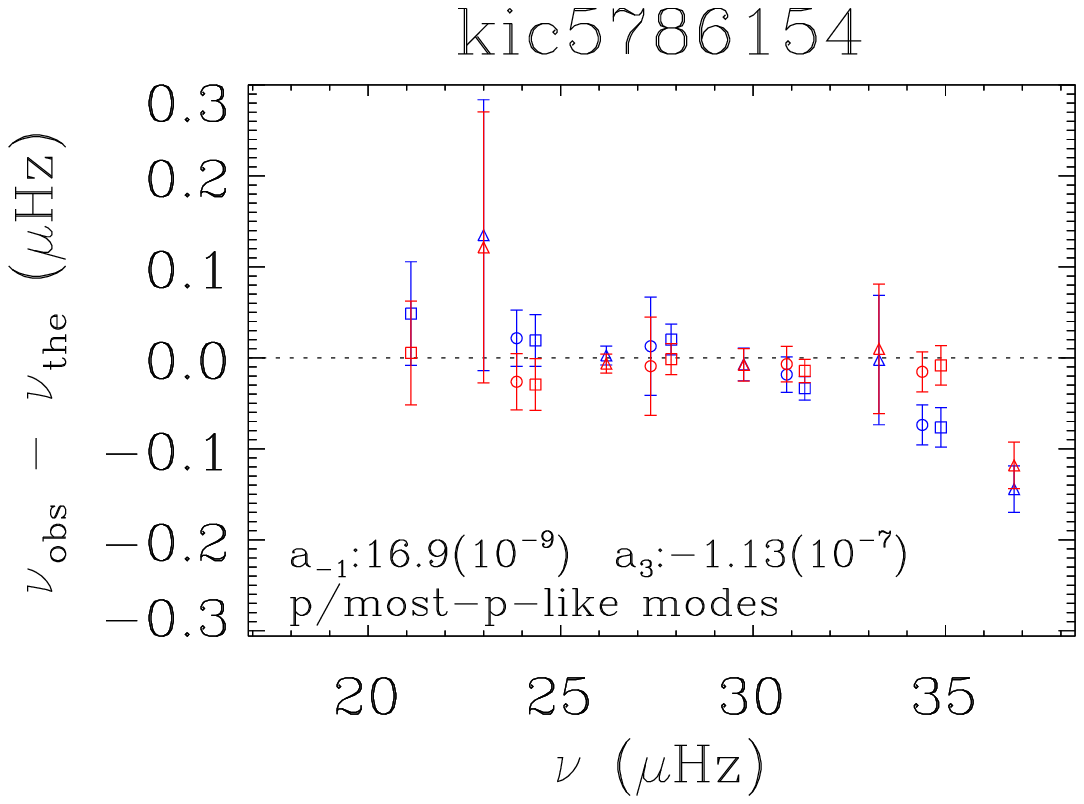}
 \includegraphics[scale=0.45]{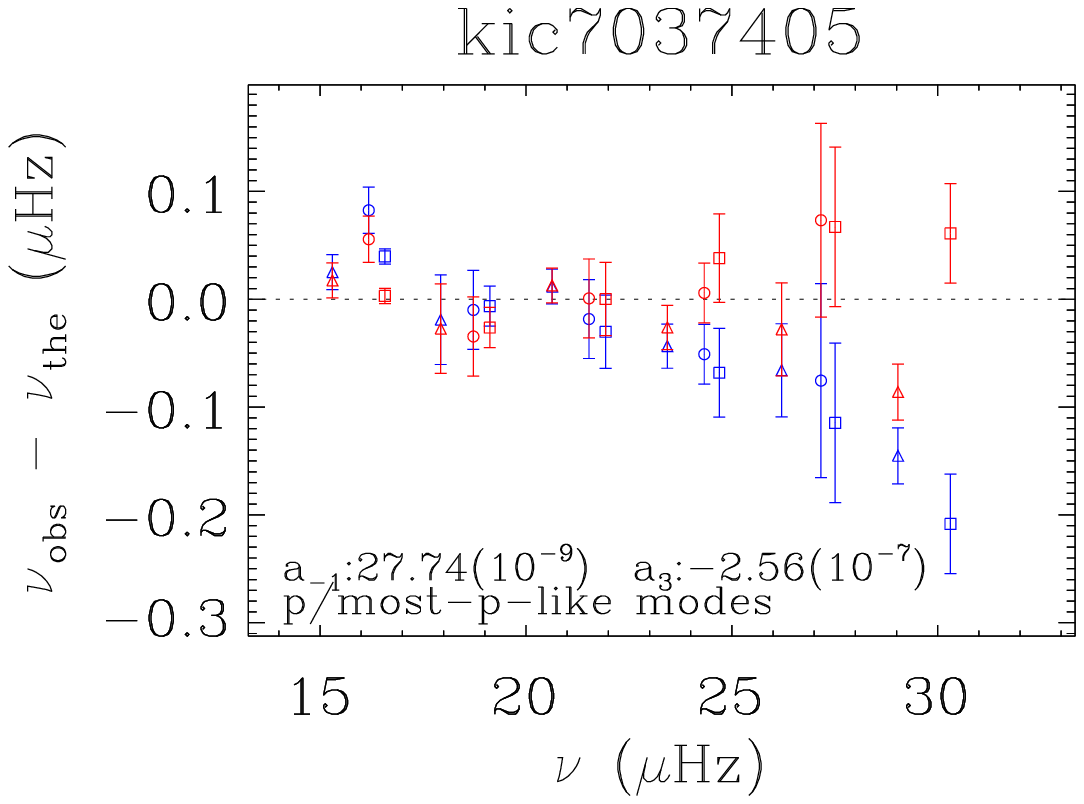}
 \includegraphics[scale=0.45]{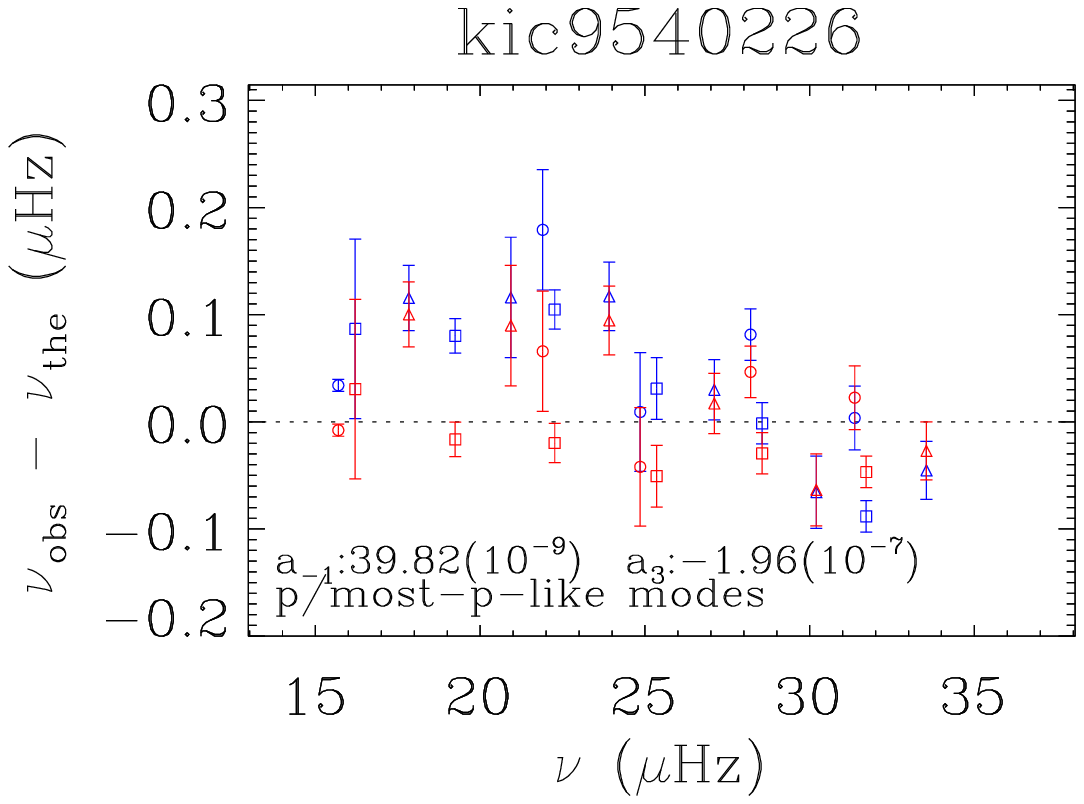}
  \caption{Frequency offsets caused by the surface term 
  of the best-fitting models of our six red giants. 
  Three stars with their $l$ = 1 mixed modes identified
are shown on the top, where upper panels include p and the most p-like modes 
for $l$ = 0 and 2, and lower panels contain p-g mixed modes for $l$ = 1. 
The other three stars with only p and the most p-like modes are shown at the bottom. 
Blue and red symbols represent the frequency offsets before and after the surface correction. 
Open squares, triangles, and circles indicate $l$ = 0, 1, and 2 modes. Error bars are
observation uncertainties. 
  }
 \label{fig:sc}
 \end{figure*}  

\subsection{The Surface Term in Evolved Stars}
 
The surface effect, which correlates with the near surface properties, 
is believed to be a function of $T_{\rm{eff}}$ and $\log g$ \citep{sonoi15}. 
The comparison between hydrodynamical simulations
and stellar models \citep{ball16} suggested an increasing 
surface term for dwarfs
from the spectral type of K5 to F3.  
\citet{Trampedach17} estimated the frequency shifts in
various types of stars with a 
grid of convection simulations and also gave
a clear correlation between the surface term and
atmospheric parameters. 

\begin{figure}
\includegraphics[scale=.6]{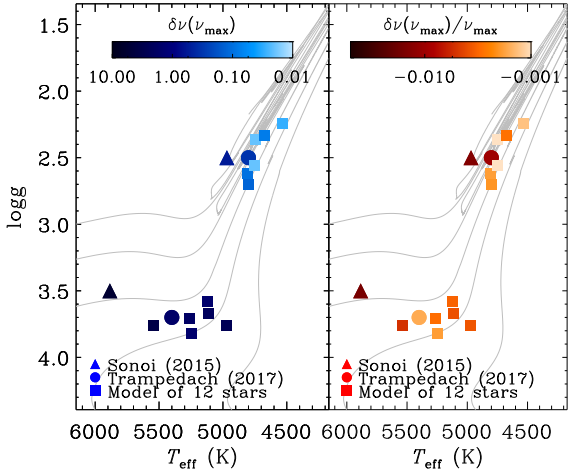}
 \caption{Absolute (left) and relatively (right) frequency shifts across the HR diagram.
          The red giants studied in this work and six low-luminosity giants 
          from \citet{ball17} are shown by squares. Filled triangles and
          circles are simulating predictions
          from \citet{sonoi15} and \citet{Trampedach17}. 
          Frequency shifts are colour scaled according
          to the colour bar shown on the top.       
         } \label{fig:sc_hr}
\end{figure}

The frequency offsets caused by the surface term 
in the best-fitting models of our six red giants are given in Fig. \ref{fig:sc}. 
All these stars resulted in the offsets in order of
0.1 $\mu$Hz at $\nu_{\rm{max}}$.
We compare our results with previous studies in
Fig. \ref{fig:sc_hr}, which gives the distribution of 
absolute ($\delta \nu$) and relative ($\delta \nu$/$\nu_{\rm{max}}$) 
frequency offsets as function of the location in the HR diagram.
In order to compare with previous results, we also included the six low luminosity red giants 
studied by \citet{ball17} as well as the simulation values 
given by \citet{sonoi15} and \citet{Trampedach17}.
Modelling results between low and high
luminosity red giants show an apparent reduction in both absolute and relative offsets 
when stars become more evolved. The decreasing $\delta \nu$ 
on the red giant branch from modelling agrees with the trend predicted by 
simulation. However, the fractional change ($\delta \nu$/$\nu_{\rm{max}}$)
does not follow the simulation results.  

\begin{figure}
\includegraphics[width=\columnwidth]{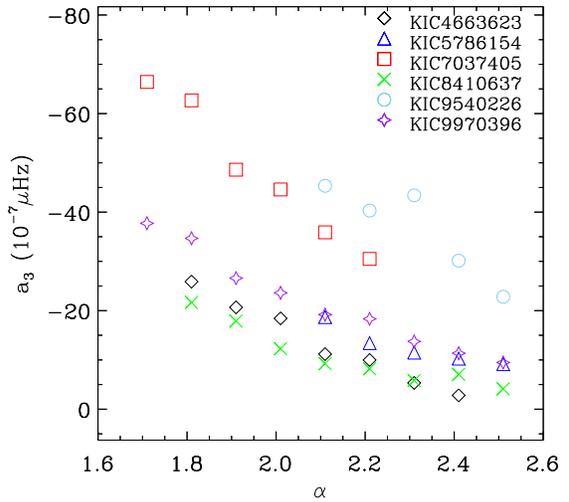}
 \caption{The correlation between the mixing-length parameter
and the surface term. The coefficient of the cubic term ($a_{3}$)
is from the best-fitting model for each star. 
 } \label{fig:alpha_a3}
\end{figure}

Apart from the improper modelling for the surface layers,   
the uncertainty of the mixing-length parameter also 
affects the surface term.
Fig. \ref{fig:alpha_a3} shows the correlation between the 
mixing-length parameter and the surface-correction coefficient of 
the cubic term ($a_{3}$). The surface term, as expected, shows
a strong dependence on the input $\alpha$ of the model.
It varies by a factor ranging from 1.5 to 2 
for different stars when the mixing length parameter
changing from 1.9 (the solar value) to 2.2 (the red giants value). 
This is to say, if $\alpha$ can not be handled well, 
the surface term will come with significant scatter.
Hence the calibrated mixing-length parameter for the six red giants
also provide a good constraint on the surface term coefficients.
We state here that the degeneracy between $\alpha$ and $a_3$ 
does not influence our results in Fig. 7, as we have marginalised over all other 
variables when deriving the posteriors, so these degeneracies are 
captured in the uncertainties.

\begin{figure}
\includegraphics[width=\columnwidth] {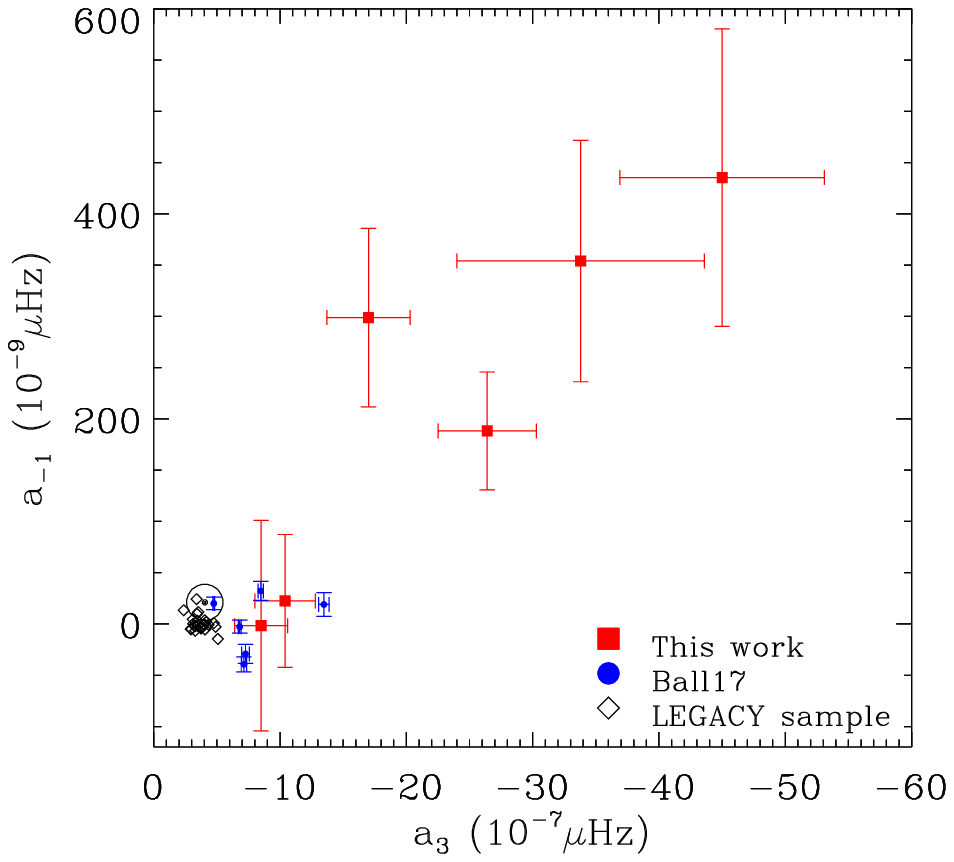}
 \caption{The correlation of two free coefficients ($a_{-1}$ and $a_{3}$)
         in surface correction expression (Eq. 6). Filled squares indicate the results of our
         red giants. Open circles represent the 6 low-luminosity giants
         from \citet{ball17}. Open diamonds are 27 dwarfs from the LEGACY sample.  
         Solar value is represented by $\odot$. 
         } \label{fig:a1_a3}
\end{figure}

A further comparison of the coefficients of stars
at different evolutionary stages is given in Fig. \ref{fig:a1_a3}. 
Besides the low-luminosity giants from \citet{ball17} and the high-luminosity giants
in this work, we also selected 27 main-sequence dwarfs 
from the LEGACY sample. These stars 
cover a range from 0.75 to 1.15$M_{\odot}$, $T_{\rm{eff}}$ = 5500 - 6200K 
and $\log g$ = 4.2 -- 4.5. 
The comparison shows a significantly increase of $a_{-1}$ and $a_3$ with stellar
evolution. We note that the increase is because of the growth 
of mode inertia, not indicating an increasing surface term with stellar evolution.  
On the contrary, the fractional frequency offsets ($\delta \nu$/$\nu_{\rm{max}}$)
decrease with stellar evolution.
Moreover, a good linear relation between $a_{-1}$ and $a_{3}$
shows in our six red giants: $a_{3}$ $\sim$ -10 $a_{-1}$. 
This agrees with the simulation result from \citet{Trampedach17},
who found a stable ratio between the inverse and cubic terms for red giants 
with similar $T_{\rm{eff}}$ and $\log g$. 

\section{Conclusions}
The mixing-length parameter in stellar models, 
responsible for the strength of energy transport in convective regions, 
is important for the accuracy of the theoretical model. 
Moreover, the surface term affects the accuracy of asteroseismic modes and
brings additional uncertainties in seismic products.
In this work, six oscillating red giants in eclipsing binaries 
were used to calibrate these parameters for evolved stars.
Our main conclusions from the results
are summarised as follows:

\begin{enumerate}
\item The average mixing-length parameter of the six red giants is $\sim$ 1.14 $\pm$ 0.07 times
the calibrated solar value, which is similar with the previous results 
based on the APOGEE sample \citep{Tayar17}.

\item Our calibrated $\alpha$ is about 16\% higher than the value
given by the 3-D hydrodynamical simulations, possibly indicating that the 1-D stellar model does not model the near-surface layers appropriately for the red giants with the input physics in this work.

\item The surface term was found to affect the mixed modes indirectly. 
Its effect on acoustic waves changes the frequency range 
where the p-g coupling happens.

\item For our six red giants, established surface-correction methods fail to fix the surface effects in g-dominated modes, which cause a non-physical reordering in mixed modes.

\item The surface term correlates with surface 
properties ($T_{\rm{eff}}$ and $\log g$) as well as the mixing-length parameter.
The frequency offset decreases with stellar evolution on the red giant branch.

\item The two coefficients ($a_{-1}$ and $a_{3}$) in the surface-correction expression (Eq. 6) 
significantly increase with stellar evolution due to the growth
of mode inertia. They also show a linear correlation in the six red giants. 
\end{enumerate}

The calibrated results of the mixing-length parameter and the surface term on the six red giants can be additional references, along with the Sun, for further studies of red giants. 
The results can improve the accuracy of he theoretical models of stellar evolution and stellar oscillation by narrowing down the ranges of free parameters. 
High-precision measurements of the stellar atmosphere and other key parameters, such as the mass and radius, are required to further constrain theoretical models.

\section*{Acknowledgements}
We gratefully acknowledge the entire $Kepler$ team and 
everyone involved in the $Kepler$ mission for making this 
paper possible. Funding for the $Kepler$ Mission is provided 
by NASA's Science Mission Directorate. 
The authors thank J\o{}rgen Christensen-Dalsgaard for fruitful discussions 
of the asteroseismic surface term and the oscillation code (ADIPLS), 
Sanjib Sharma for the fruitful discussion of the Bayesian method, and 
Karsten Brogaard for the helpful comments during the KASOC review. 
This work is supported by an Australian Research Council DP grant DP150104667 
awarded to JBH and TRB, the Danish National Research Foundation (Grant DNRF106), 
grants 11503039, 11427901, 11273007, and 10933002 from the National Natural Science Foundation of China, and the Deutscher Akademischer Austauschdienst (DAAD) through the Go8 Australia-Germany Joint Research Co-operation Scheme.   
WHB acknowledges funding from the UK Science and Technology Facilities Council (STFC).

\clearpage

\appendix

\begin{figure*}
\includegraphics[width=\textwidth]{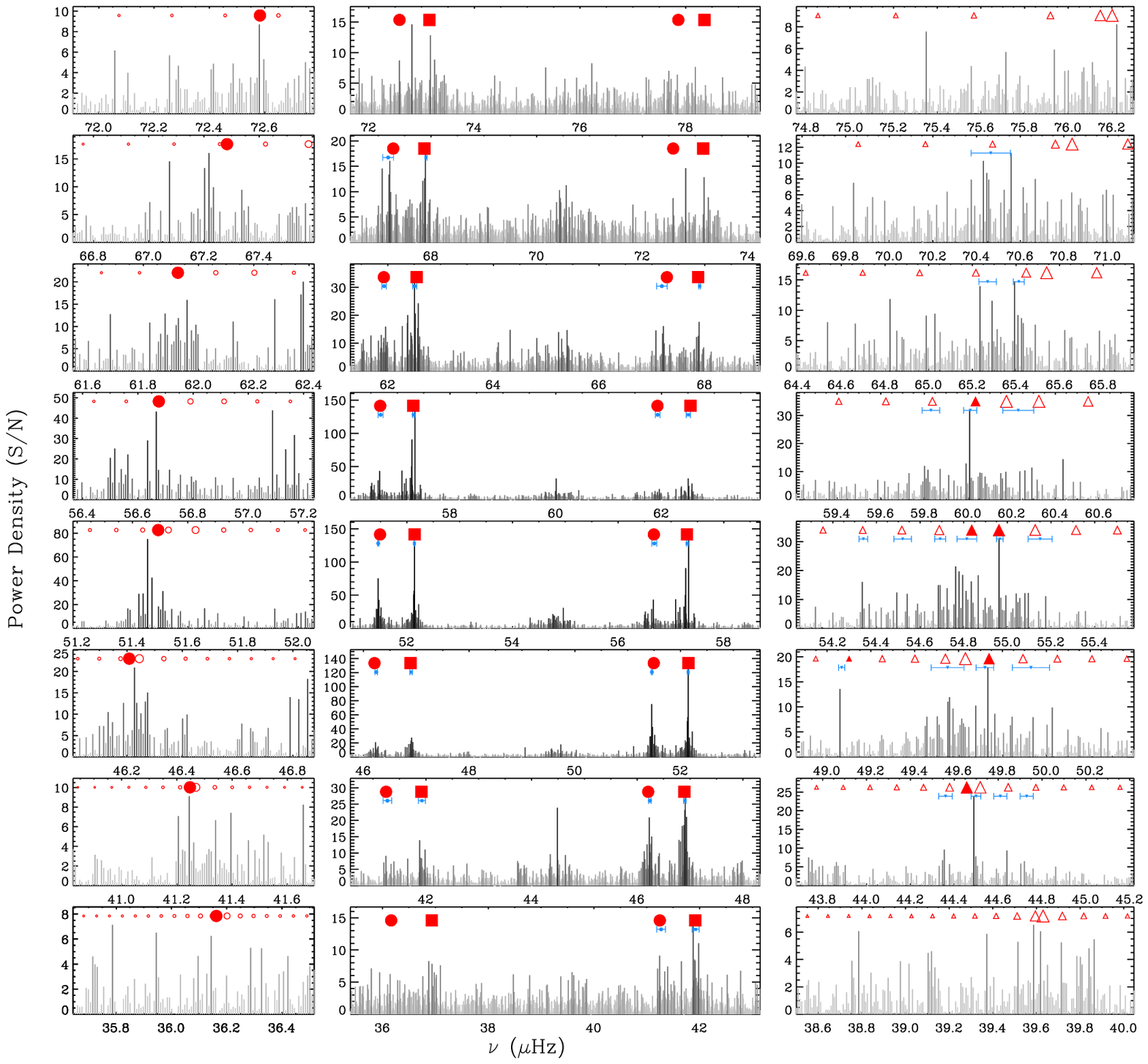}
 \caption{The peak-bagging process of individual mixed modes
for KIC 4663623. 
The whole power spectrum is separated into eight radial-mode orders
as shown in the middle. Close inspections of $l$ = 2 and 1 modes are plotted on the left and right sides. 
The colour code was set as same as that in Figure 1, 
indicating the $\mathcal{P}_{\rm{signal}}$
of each frequency bin.
Red Symbols plotted on the top are theoretical
frequencies of the best fitting model. Squares, circles and
triangles in the middle represent the p and most-p-like modes for $l$ = 0, 1 and 2.
Filled symbols are the modes for picking the fitting models, and 
open symbols are those for guiding the identification of other mixed modes.
Circles and triangles on the left and right indicate all theoretical mixed modes in each frequency bin. And their symbol size is scaled 
with $1/I^{2}$ ($I$ is mode inertia) by reference to that of the
most p-like mode in each degree and order.
Larger size indicates the mode to be more p-like and less in inertia.     
Small blue symbols represent identified observed frequencies.
} \label{fig:4663623}
\end{figure*}

\clearpage

\begin{figure}
\includegraphics[width=\columnwidth]{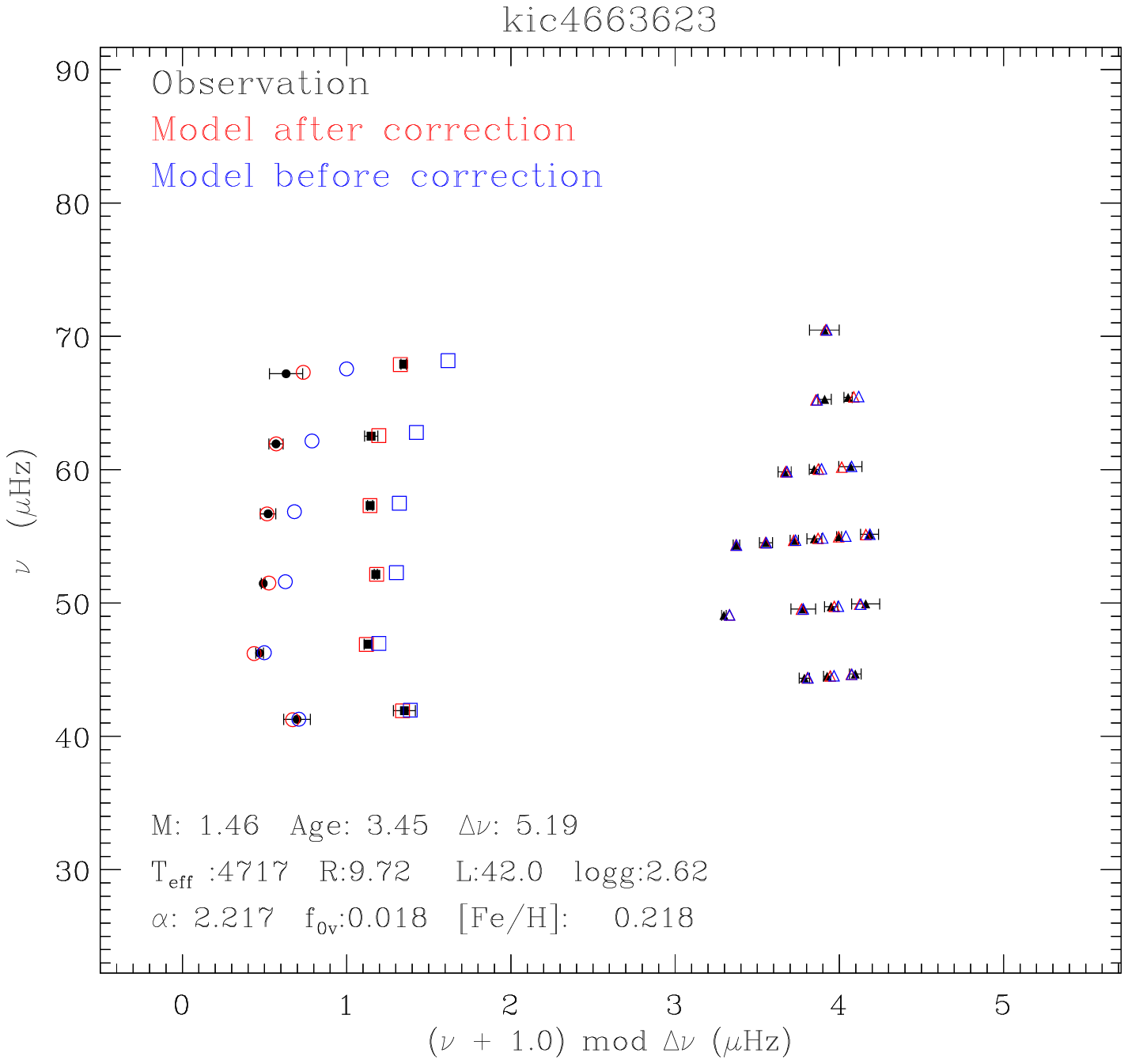}
 \caption{Observed and theoretical \'echelle diagram of KIC4663623. 
 Black symbols are observed modes. Blue and red symbols
 indicate model frequencies before and after the surface correction. 
 } \label{fig:4663623ed}
\end{figure}

\begin{table}
 \centering
  \caption{Identified Oscillation Frequencies for Star KIC4663623}
  \begin{tabular}{ccc}
  \hline
$l$ & $\nu$ [$\mu$Hz] & $\sigma$ [$\mu$Hz]\\
  \hline
 \multicolumn{3}{c}{p and the most-p-like modes}\\
   \hline
     2&     41.271  &    0.081 \\    
     0&     41.924  &    0.066 \\    
     2&     46.240  &    0.024 \\     
     0&     46.900  &    0.022 \\     
     2&     51.458  &    0.011 \\     
     0&     52.144  &    0.009 \\    
     2&     56.684  &    0.046 \\     
     0&     57.305  &    0.015 \\     
     2&     61.929  &    0.043 \\    
     0&     62.509  &    0.039 \\    
     2&     67.187  &    0.101 \\
     0&     67.903  &    0.016 \\
     0&     72.820  &    0.298 \\
        \hline
  \multicolumn{3}{c}{Individual mixed modes}\\
   \hline
     1&     44.360  &    0.031 \\
     1&     44.499  &    0.022 \\
     1&     44.610  &    0.030 \\   
     1&     44.732  &    0.028 \\   
     1&     49.067  &    0.014 \\
     1&     49.550  &    0.075 \\
     1&     49.720  &    0.041 \\
     1&     49.930  &    0.085 \\
     1&     54.340  &    0.022 \\
     1&     54.520  &    0.041 \\
     1&     54.692  &    0.025 \\
     1&     54.814  &    0.045 \\
     1&     54.965  &    0.015 \\
     1&     55.150  &    0.055 \\
     1&     59.830  &    0.043 \\
     1&     60.010  &    0.028 \\
     1&     60.230  &    0.071 \\
     1&     65.270  &    0.042 \\
     1&     65.413  &    0.025 \\
     1&     70.465  &    0.090 \\
  \hline
\end{tabular}
\end{table}
\clearpage

\begin{figure*}
\includegraphics[width=\textwidth]{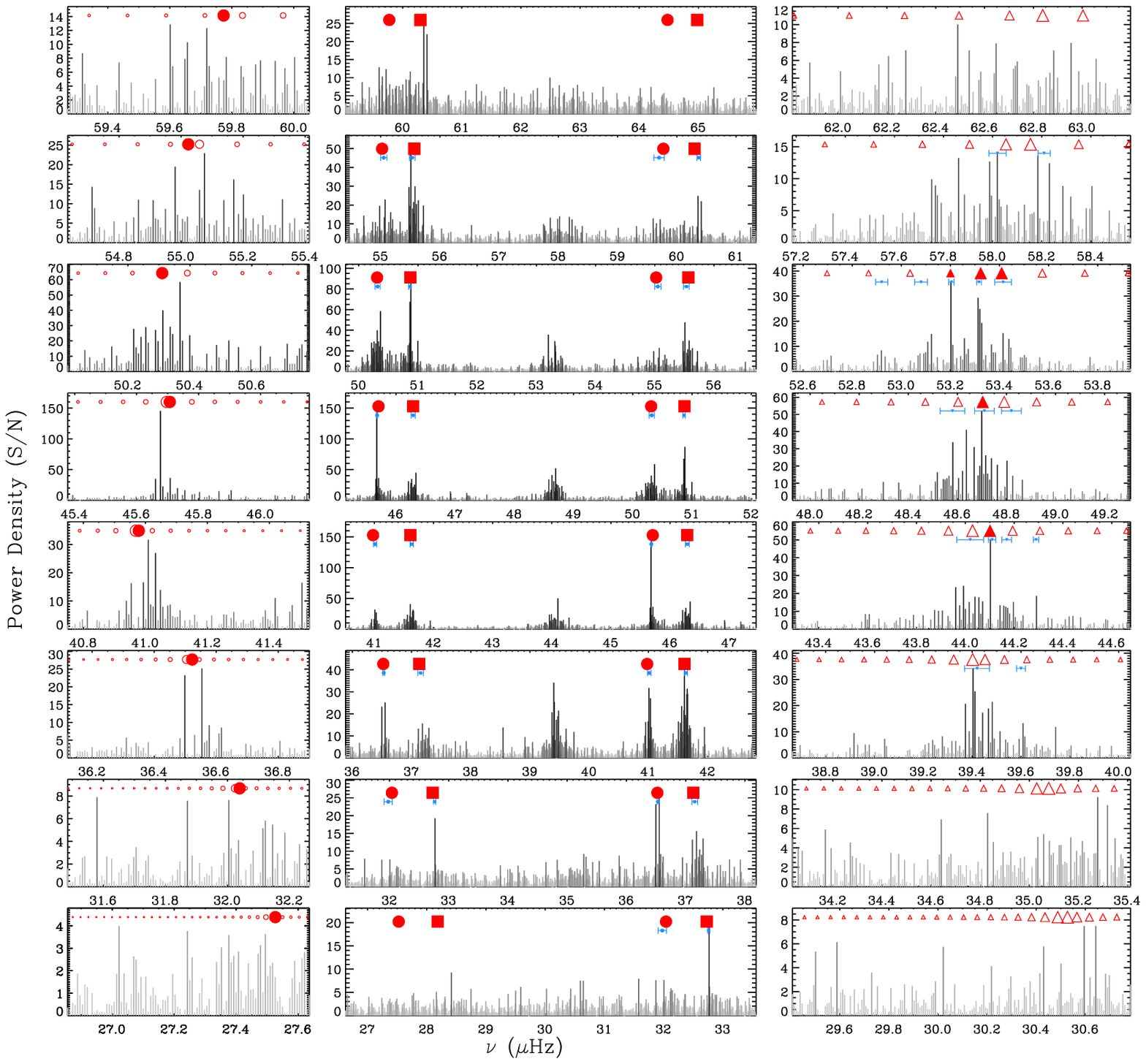}
 \caption{Same as Figure~1, but for KIC8410637.} \label{fig:8410637}
\end{figure*}
\clearpage

\begin{figure}
\includegraphics[width=\columnwidth]{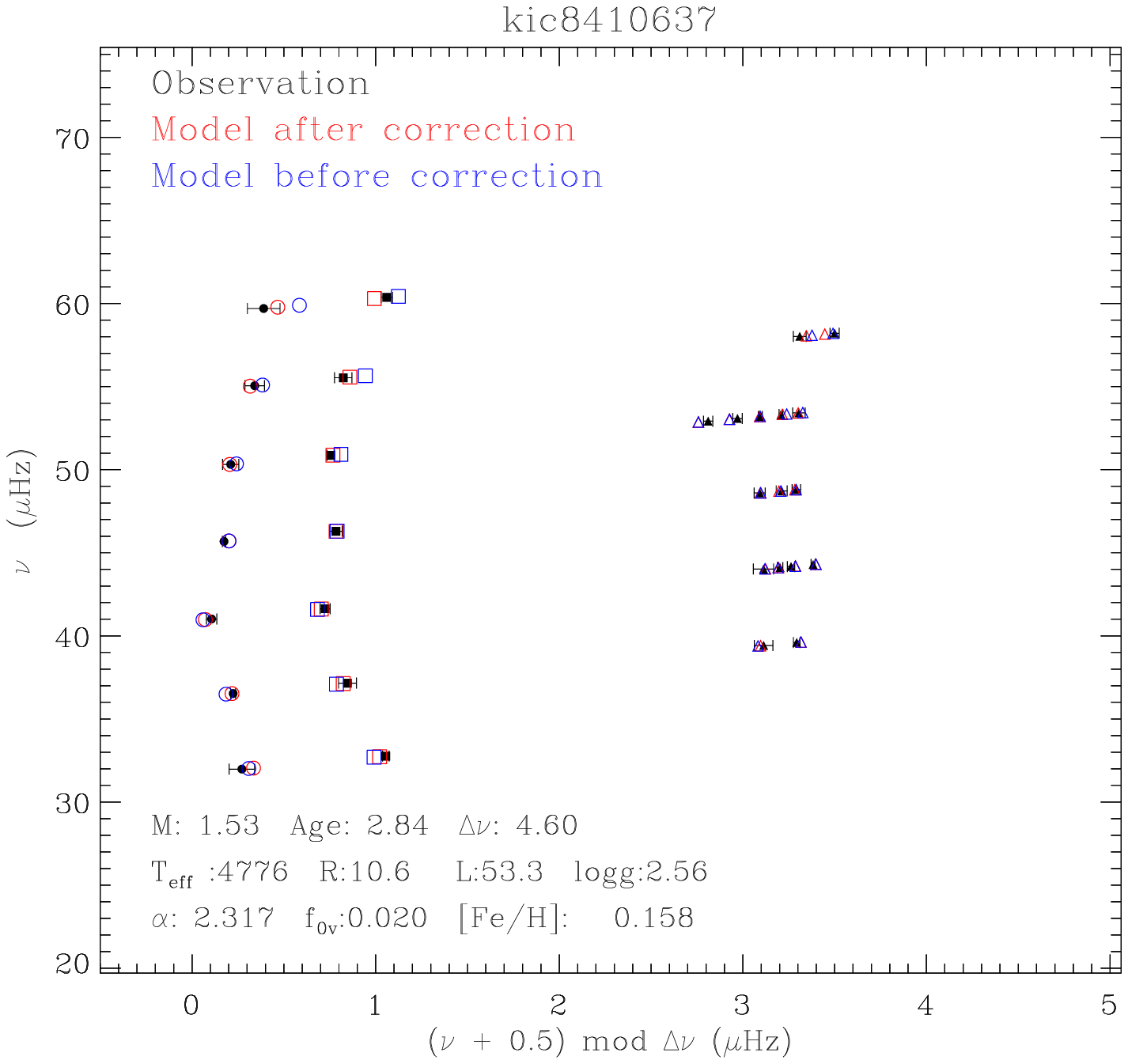}
 \caption{Same as Figure~2, but for KIC8410637.} \label{fig:8410637ed}
\end{figure}

\begin{table}
 \centering
  \caption{Identified Oscillation Frequencies for Star KIC8410637}
  \begin{tabular}{ccc}
  \hline
$l$ & $\nu$ [$\mu$Hz] & $\sigma$ [$\mu$Hz]\\
  \hline
 \multicolumn{3}{c}{p and the most-p-like modes}\\
   \hline
     2 &     41.013  &    0.028\\  
     0 &     41.632  &    0.027\\  
     2 &     45.682  &    0.009\\  
     0 &     46.291  &    0.034\\  
     2 &     50.319  &    0.043\\  
     0 &     50.864  &    0.020\\  
     2 &     55.051  &    0.052\\  
     0 &     55.533  &    0.047\\  
     2 &     59.701  &    0.088\\  
     0 &     60.371  &    0.031\\  
        \hline
  \multicolumn{3}{c}{Individual mixed modes}\\
   \hline
     1 &     39.420  &    0.051\\  
     1 &     39.600  &    0.018\\  
     1 &     44.020  &    0.055\\  
     1 &     44.110  &    0.015\\  
     1 &     44.170  &    0.02 \\  
     1 &     44.290  &    0.011\\
     1 &     48.583  &    0.048\\  
     1 &     48.712  &    0.043\\  
     1 &     48.818  &    0.039\\  
     1 &     52.920  &    0.025\\   
     1 &     53.080  &    0.026\\   
     1 &     53.203  &    0.011\\  
     1 &     53.316  &    0.010\\
     1 &     53.414  &    0.034\\
     1 &     58.020  &    0.035\\
     1 &     58.210  &    0.025\\
  \hline
\end{tabular}
\end{table}

\clearpage

\begin{figure*}
\includegraphics[width=\textwidth]{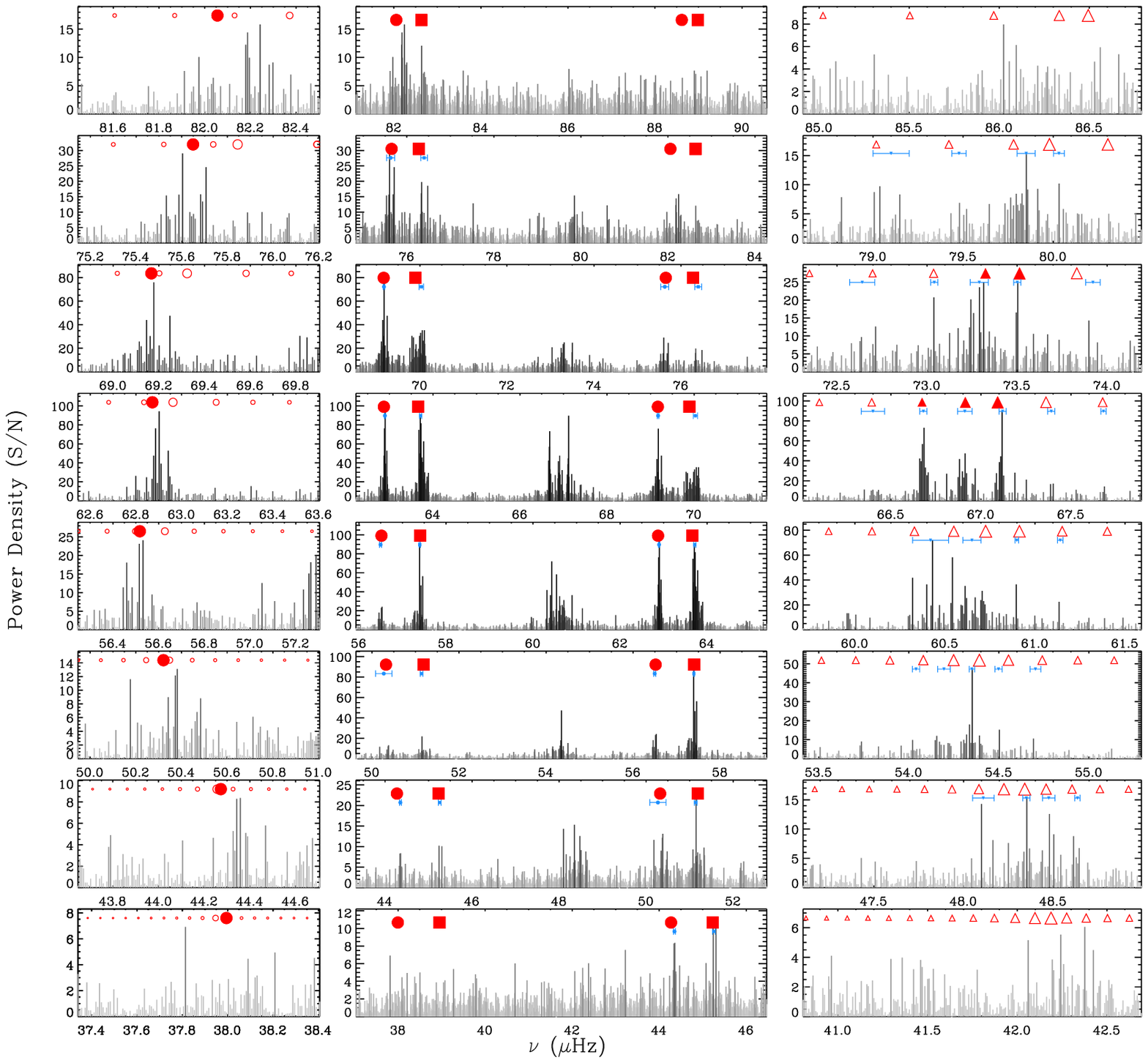}
 \caption{Same as Figure~1, but for KIC9970396.}
 \label{fig:9970396}
\end{figure*}
\clearpage

\begin{figure}
 \vspace*{300pt}
\includegraphics[width=\columnwidth]{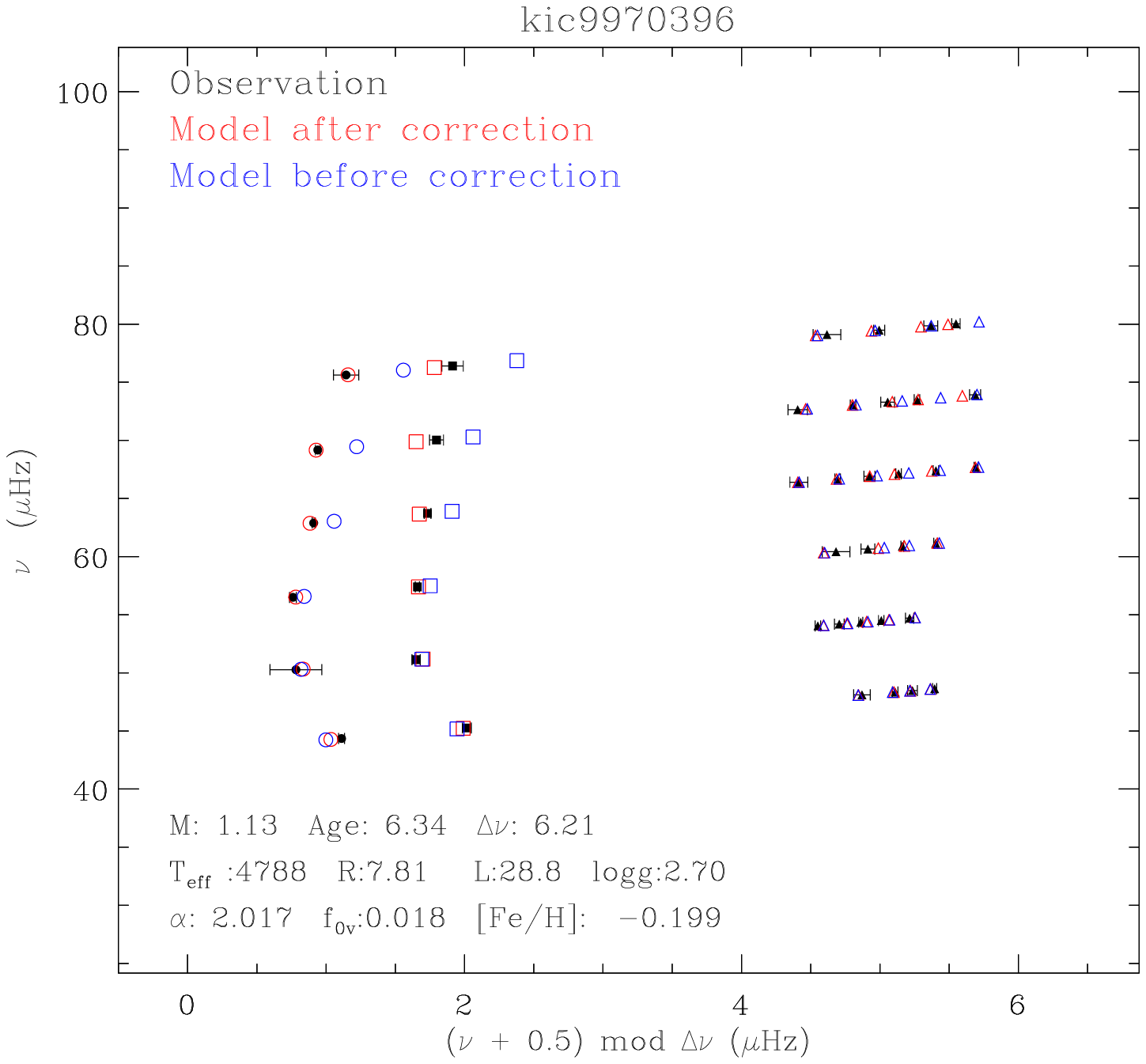}
 \caption{Same as Figure~2, but for KIC9970396.} \label{fig:9970396ed}
\end{figure}

\begin{table}
 \centering
  \caption{Identified Oscillation Frequencies for Star KIC9970396}
  \begin{tabular}{ccc}
  \hline
$l$ & $\nu$ [$\mu$Hz] & $\sigma$ [$\mu$Hz]\\
  \hline
  \multicolumn{3}{c}{p and the most-p-like modes}\\
   \hline
       2 &    44.351 &   0.021  \\  
      0 &    45.255 &   0.032  \\  
      2 &    50.270 &   0.187  \\  
      0 &    51.138 &   0.029  \\  
      2 &    56.496 &   0.023  \\  
      0 &    57.396 &   0.015  \\  
      2 &    62.895 &   0.012  \\  
      0 &    63.717 &   0.023  \\  
      2 &    69.173 &   0.017  \\  
      0 &    70.031 &   0.050  \\  
      2 &    75.626 &   0.091  \\  
      0 &    76.395 &   0.077  \\
        \hline
  \multicolumn{3}{c}{Individual mixed modes}\\
   \hline
     1  &    48.110 &   0.060  \\
     1  &    48.350 &   0.020  \\
     1  &    48.475 &   0.034  \\       
     1  &    48.635 &   0.016  \\
     1  &    54.040 &   0.018  \\               
     1  &    54.195 &   0.034  \\  
     1  &    54.350 &   0.045  \\
     1  &    54.498 &   0.020  \\
     1  &    60.375 &   0.055  \\  
     1  &    60.672 &   0.070  \\  
     1  &    60.899 &   0.012  \\
     1  &    60.899 &   0.011  \\
     1  &    61.140 &   0.015  \\
     1  &    66.400 &   0.065  \\     
     1  &    66.680 &   0.020  \\
     1  &    66.910  &  0.041  \\
     1  &    67.120 &   0.018  \\
     1  &    67.390 &   0.023  \\    
     1  &    67.680 &   0.015  \\
     1  &    72.640 &   0.070  \\     
     1  &    73.040 &   0.022  \\           
     1  &    73.290 &   0.051  \\
     1  &    73.501 &   0.019  \\     
     1  &    73.921 &   0.041  \\  
     1  &    79.080 &   0.080  \\
     1  &    79.477 &   0.040  \\
     1  &    79.800 &   0.063  \\
     1  &    80.050 &   0.018  \\              
  \hline
\end{tabular}
\end{table}

\clearpage


\begin{figure}
\includegraphics[width=\columnwidth]{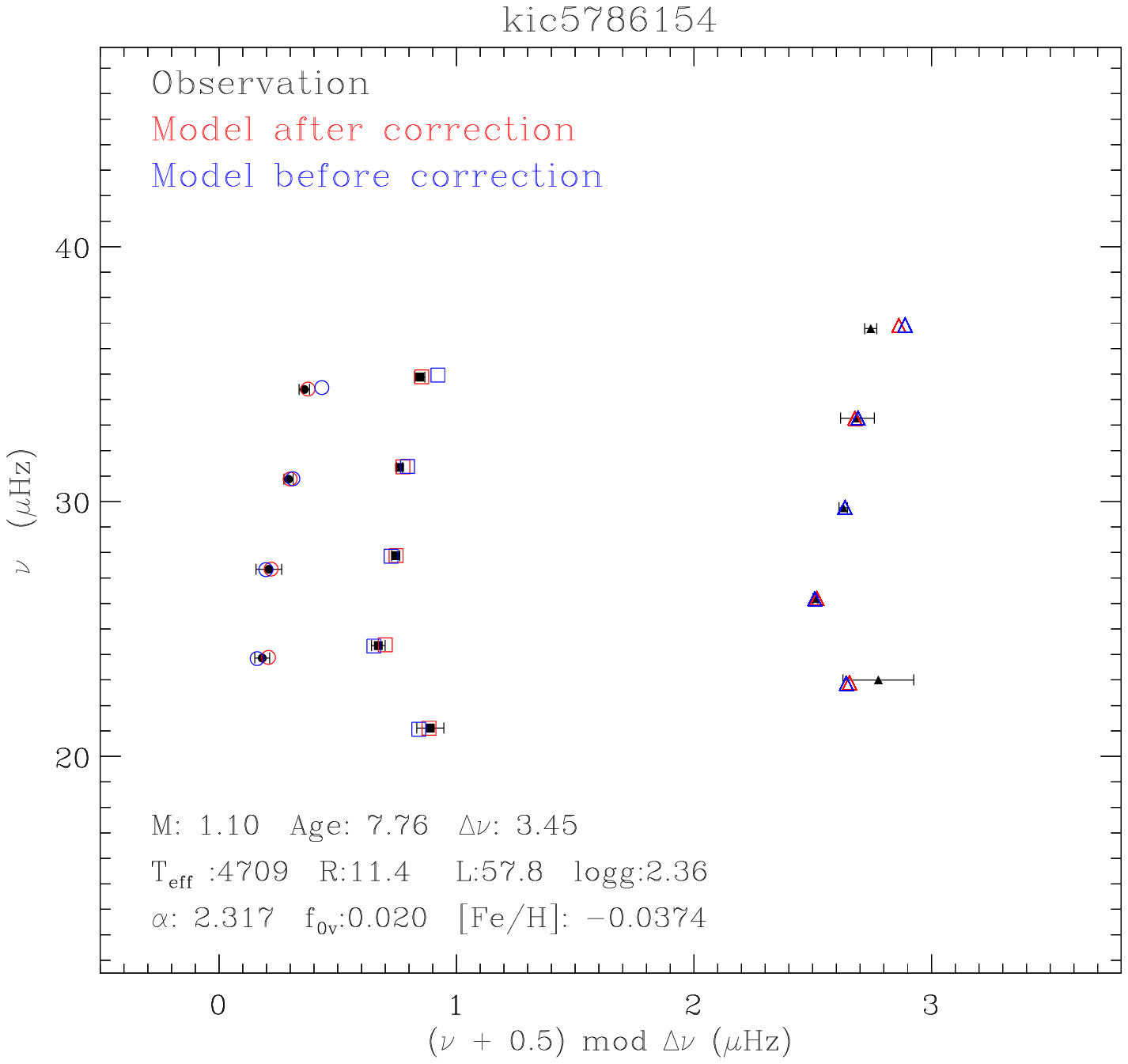}
 \caption{Same as Figure~2, but for KIC5786154.} \label{fig:5786154ed}
\end{figure}

\begin{table}
 \centering
  \caption{Identified Oscillation Frequencies for Star KIC5786154}
  \begin{tabular}{ccc}
  \hline
$l$ & $\nu$ [$\mu$Hz] & $\sigma$ [$\mu$Hz] \\
  \hline
  \multicolumn{3}{c}{p and the most-p-like modes}\\
  \hline
      0 &    21.112 &     0.056  \\ 
      1 &    22.647 &     0.048 \\
      2 &    23.858 &     0.030 \\
      0 &    24.347 &     0.028 \\
      1 &    26.187 &     0.010 \\
      2 &    27.340 &     0.054 \\
      0 &    27.874 &     0.016 \\
      1 &    29.758 &     0.017 \\
      2 &    30.876 &     0.019 \\
      0 &    31.343 &     0.012 \\ 
      1 &    33.271 &     0.071 \\
      2 &    34.396 &     0.021 \\
      0 &    34.882 &     0.021 \\
      1 &    36.781 &     0.025 \\
  \hline
\end{tabular}
\end{table}

\clearpage


\begin{figure}
\includegraphics[width=\columnwidth]{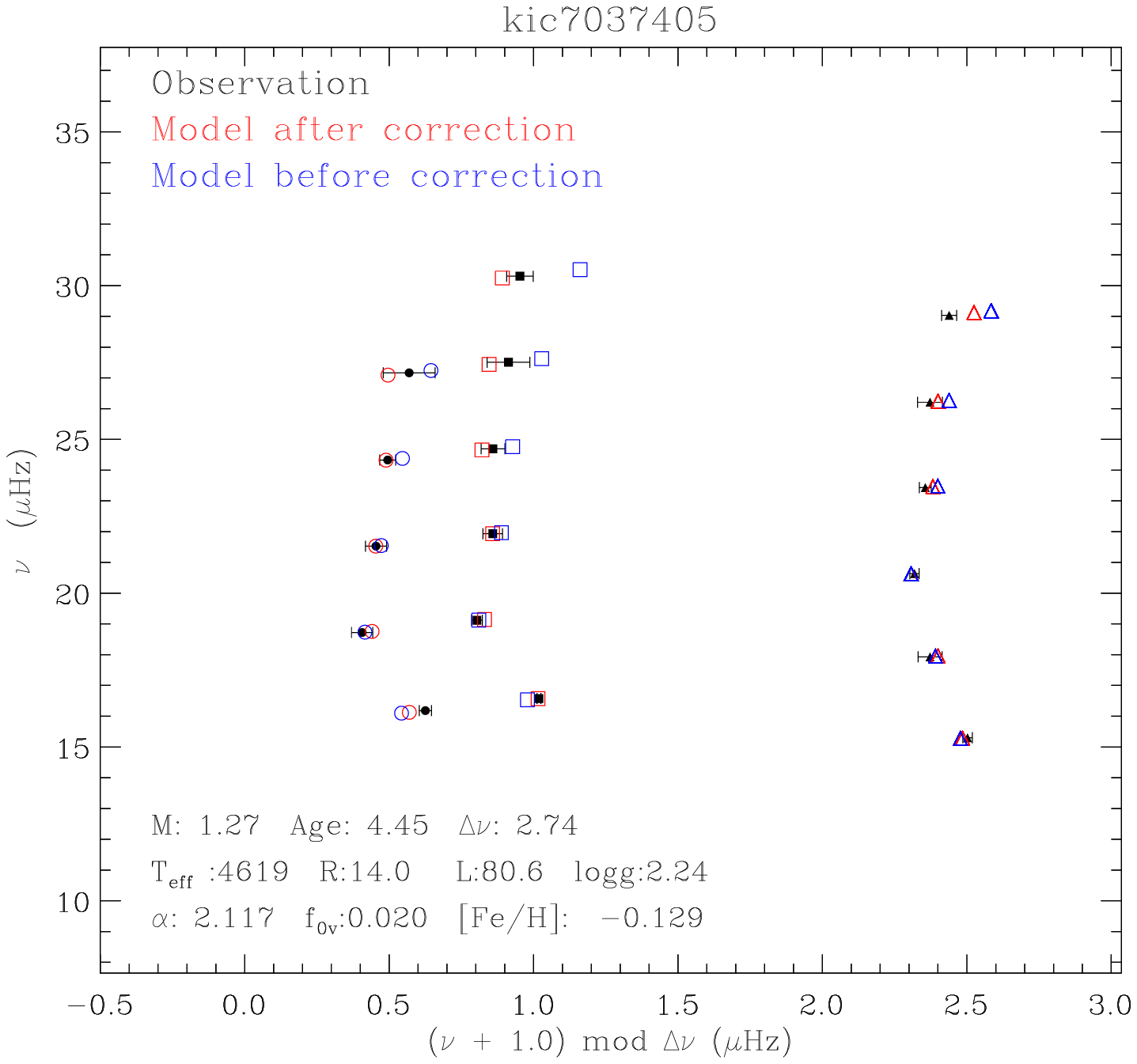}
 \caption{Same as Figure~2, but for KIC7037405.} \label{fig:7037405ed}
\end{figure}

\begin{table}
 \centering
  \caption{Identified Oscillation Frequencies for Star KIC7037405}
  \begin{tabular}{ccc}
  \hline
$l$ & $\nu$ [$\mu$Hz] & $\sigma$ [$\mu$Hz]\\
  \hline
    \multicolumn{3}{c}{p and the most-p-like modes}\\
  \hline
      1&    15.310  &    0.016\\     
      2&    16.182  &    0.021\\
      0&    16.575  &    0.007\\
      1&    17.929  &    0.041\\
      2&    18.723  &    0.036\\
      0&    19.120  &    0.018\\
      1&    20.635  &    0.016\\
      2&    21.530  &    0.036\\
      0&    21.935  &    0.034\\
      1&    23.432  &    0.020\\
      2&    24.331  &    0.027\\
      0&    24.695  &    0.041\\
      1&    26.208  &    0.043\\
      2&    27.164  &    0.089\\      
      0&    27.508  &    0.074\\
      1&    29.034  &    0.025\\
      0&    30.307  &    0.046\\
  \hline
\end{tabular}
\end{table}

\clearpage


\begin{figure}
\includegraphics[width=\columnwidth]{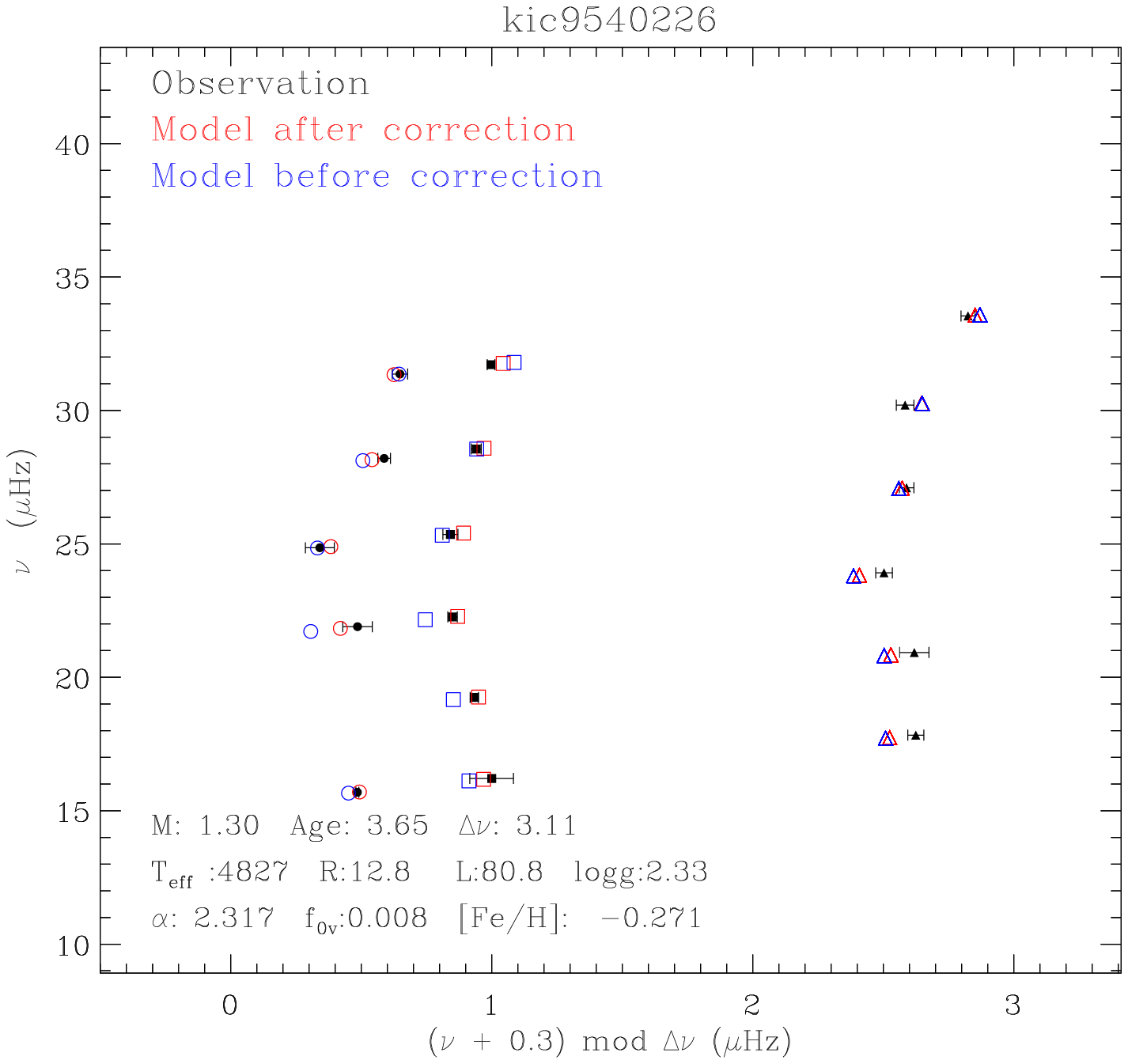}
 \caption{Same as Figure~2, but for KIC9940226.} \label{fig:9940226ed}
\end{figure}

\begin{table}
 \centering
  \caption{Identified Oscillation Frequencies for Star KIC9540226}
  \begin{tabular}{ccc}
  \hline
$l$ & $\nu$ [$\mu$Hz] & $\sigma$ [$\mu$Hz]\\
  \hline
    \multicolumn{3}{c}{p and the most-p-like modes}\\
  \hline
2&  15.694 &    0.005 \\
0&  16.208 &    0.083 \\ 
1&  17.833 &    0.030 \\ 
0&  19.244 &    0.016 \\ 
1&  20.929 &    0.056 \\
2&  21.898 &    0.056 \\  
0&  22.262 &    0.018 \\ 
3&  22.819 &    0.016 \\ 
1&  23.915 &    0.032 \\ 
2&  24.785 &    0.030 \\ 
0&  25.355 &    0.028 \\ 
1&  27.103 &    0.028 \\
2&  28.203 &    0.024 \\ 
0&  28.556 &    0.019 \\ 
1&  30.199 &    0.033 \\ 
2&  31.366 &    0.029 \\
0&  31.714 &    0.014 \\
1&  33.542 &    0.027 \\
  \hline
\end{tabular}
\end{table}

\bsp

\label{lastpage}

\end{document}